\documentclass[10pt,journal,compsoc]{IEEEtran}

\usepackage{colortbl}
\usepackage{pgfplots}
\usepackage{pgfplotstable}

\usepackage{tikz}
\usetikzlibrary{patterns}

\usepackage{listings}
\usepackage{flushend}
\usepackage{mdwlist}

\ifCLASSOPTIONcompsoc
  \usepackage[nocompress]{cite}
\else
  \usepackage{cite}
\fi

\usepackage{enumitem}
\setlist{nosep}

\usepackage{amsthm}
\theoremstyle{definition}
\newtheorem{cor}{Assumption}

\usepackage{graphicx}
\graphicspath{{./pdf_pic/}}

\usepackage[pdftex]{hyperref}

\usepackage{array}

\ifCLASSOPTIONcompsoc
  \usepackage[caption=false,font=footnotesize,labelfont=sf,textfont=sf]{subfig}
\else
  \usepackage[caption=false,font=footnotesize]{subfig}
\fi

\usepackage{multirow}
\usepackage{url}

\newcommand{\tc}[1]{\multicolumn{1}{c|}{#1}}

\usepackage[cmex10]{amsmath}

\usepackage{url}

\begin{document}
%

\title{Dissecting GPU Memory Hierarchy through Microbenchmarking}

\author{Xinxin~Mei,
    Xiaowen~Chu,~\IEEEmembership{Senior Member,~IEEE}

\IEEEcompsocitemizethanks{\IEEEcompsocthanksitem Xinxin Mei and Xiaowen Chu are with the Department
of Computer Science, Hong Kong Baptist University. Xiaowen Chu is also with HKBU Institute of Research and Continuing Education. \protect\\E-mail: \{xxmei, chxw\}@comp.hkbu.edu.hk
\IEEEcompsocthanksitem Our source code and experimental data are publicly available at: \protect\\ \href{http://www.comp.hkbu.edu.hk/\~chxw/gpu_benchmark.html}{http://www.comp.hkbu.edu.hk/$\sim$chxw/gpu\_benchmark.html}.}
}

\IEEEtitleabstractindextext{%
\begin{abstract}

Memory access efficiency is a key factor in fully utilizing the computational power of graphics processing units (GPUs). However, many details of the GPU memory hierarchy are not released by GPU vendors. In this paper, we propose a novel fine-grained microbenchmarking approach and apply it to three generations of NVIDIA GPUs, namely Fermi, Kepler and Maxwell, to expose the previously unknown characteristics of their memory hierarchies. Specifically, we investigate the structures of different GPU cache systems, such as the data cache, the texture cache and the translation look-aside buffer (TLB). We also investigate the throughput and access latency of GPU global memory and shared memory. Our microbenchmark results offer a better understanding of the mysterious GPU memory hierarchy, which will facilitate the software optimization and modelling of GPU architectures. To the best of our knowledge, this is the first study to reveal the cache properties of Kepler and Maxwell GPUs, and the superiority of Maxwell in shared memory performance under bank conflict.

\end{abstract}

}

\maketitle
\IEEEdisplaynontitleabstractindextext

\IEEEpeerreviewmaketitle

\section{Introduction}

\IEEEPARstart{T}he past decade has witnessed a boom in the development of general-purpose graphics processing units (GPGPUs). These GPUs are embedded with hundreds to thousands of arithmetic processing units on one die and have tremendous computing power. They are one of the most successful types of many-core parallel hardware and are deployed in a great variety of scientific and commercial applications. The prospect of more thorough and broader applications is very promising \cite{nickolls2010gpu,Hwu20142574,keckler2011gpus,zhao2014g,li2014accelerating,ryoo2008optimization}. However, their realistic performance is often limited by the huge performance gap between the processors and the GPU memory system. For example, NVIDIA's GTX980 has a raw computational power of 4,612 GFlop/s, but its theoretical memory bandwidth is only 224 GB/s \cite{MaxwellWhitepaper}. The realistic memory throughput is even lower. The memory bottleneck remains a significant challenge for these parallel computing chips \cite{Hwu20142574,keckler2011gpus}. The GPU memory hierarchy is rather complex, and includes the GPU-unique shared, texture and constant memory. According to the literature, appropriate leverage of GPU memory hierarchies can provide significant performance improvements \cite{micikevicius20093d,zhao2014g,li2014accelerating,ryoo2008optimization}. For example, on GTX780, the memory-bound G-BLASTN achieves an overall 14.8x speedup compared with the sequential NCBI-BLAST by coordinating the use of GPU texture and shared memory \cite{zhao2014g}. On GTX980, the performance of a naive compute-bound matrix multiplication kernel without memory optimization is only 148 GFlop/s, that of a kernel with clever application of shared memory is 598 GFlop/s, and that of a kernel with extremely efficient optimization of memory is as high as 1,225 GFlop/s \cite{matrixmul_SDK,matrixmulCUBLAS_SDK}. Hence, it is vital to expose, exploit and optimize GPU memory hierarchies.

NVIDIA has launched three generations of GPUs since 2009, codenamed as Fermi, Kepler and Maxwell, with compute capabilities of 2.x, 3.x and 5.x, respectively. Compared with its former 1.x hardware, NVIDIA has devoted much effort to improving GPU memory efficiency, yet the memory bottleneck is still a primary limitation \cite{fermiwhitepaper,keplerwhitepaper,keplertuningguide,MaxwellWhitepaper,MaxwellTuning,cudacprogrammingguide,cudabestguide}. Because NVIDIA provides very limited information on its GPU memory systems, many of their details remain unknown to the public. Existing work on the disclosure of GPU memory hierarchy is generally conducted using third-party benchmarks \cite{volkov2008benchmarking,papadopoulou2009micro,wong2010demystifying,zhang2011quantitative,
baghsorkhi2012efficient,meltzer2013micro}. Most of them are based on devices with a compute capability of 1.x \cite{volkov2008benchmarking,papadopoulou2009micro,wong2010demystifying,zhang2011quantitative}. Recent explorations of Fermi architecture focus on a part of the memory system \cite{baghsorkhi2012efficient,meltzer2013micro}. To the best of our knowledge, there are no state-of-the-art works on the recent Kepler and Maxwell architectures. Furthermore, the above benchmark studies on GPU cache structure are based on a method that was developed for early CPU platforms \cite{saavedra1992cpu, saavedra1995measuring} with a simple memory hierarchy. As memory designs have become more sophisticated, this method has become out of date and inappropriate for current generations of GPU hardware \cite{mei2014benchmarking}.

In this paper, we investigate the GPU memory hierarchy of three recent generations of NVIDIA GPUs: Fermi, Kepler and Maxwell. We investigate them using a series of microbenchmarks targeting their cache mechanism, memory throughput, and memory latency. In particular, we propose a fine-grained pointer chasing (P-chase) microbenchmark, which reveals that many of the characteristics of a GPU cache differ from those of a CPU. All our experimental results are based on many rounds of experiments and are reproducible. Our work illuminates the currently mysterious architecture of GPU memory. In addition, by comparing the properties of three generations of GPU memory hierarchy, we can clearly perceive the evolution of GPU memory designs. The Kepler device is designed to maximize compute performance by aggressively integrating many emerging technologies, whereas the latest Maxwell device is more conservative and aims at energy efficiency rather than pure compute performance.

We highlight the contributions of our work as follows.
\begin{enumerate}
  \item We propose a novel fine-grained P-chase microbenchmark to explore the unknown GPU cache parameters. Our results indicate that GPUs have many features that differ from those of traditional CPUs. We discover the unequal sets of L2 translation look-aside buffer (TLB), the 2D spatial locality optimized set-associative mapping of the texture L1 cache, and the non-traditional replacement policy of the L1 data cache.
  \item We quantitatively benchmark the throughput and access latency of a GPU's global and shared memory. We study the various factors that influence the memory throughput, and the effect of the shared memory bank conflict on the memory access latency. For the first time, we verify that Maxwell is highly optimized to avoid long latency under shared memory bank conflict.
  \item Our work provides comprehensive and up-to-date information on the GPU memory hierarchy. Our microbenchmarks cover the architecture, throughput and latency of recent generations of GPUs. To the best of our knowledge, this paper is the first to study the new features of the Kepler and Maxwell GPUs.
\end{enumerate}

The remainder of this paper is organized as follows. Section 2 summarizes the related work and Section 3 gives an overview of GPU memory hierarchy. Section 4 introduces the fine-grained P-chase microbenchmark and how we apply it to dissect the GPU cache micro-architectures. Section 5 presents our study on the effective global memory throughput and memory access latencies under different access patterns. Section 6 investigates the shared memory in terms of latency, throughput, and the impact of bank conflict. We conclude our findings in Section 7.

\section{Related Work}

Many studies have investigated GPU memory system, some of which have confirmed that the performances of many GPU computing applications are limited by the memory bottleneck \cite{ryoo2008optimization,Hwu20142574,keckler2011gpus,lal2014gpgpu}. Using characterization approaches, several studies located the causes of low memory throughput to relative memory spaces \cite{ryoo2008optimization,lal2014gpgpu,jia2012characterizing}. Some designed a number of data mapping/memory management algorithms with the aim to improve memory access efficiency \cite{jia2012characterizing,xie2013efficient,jang2011exploiting,Che2011Dymaxion,sung2010data}. Recently, Li et al. proposed a locality monitoring mechanism to better utilize the L1 data cache for higher performance \cite{li2015}. All of these studies have contributed to the field and inspired our work.

\begin{table}
\renewcommand{\arraystretch}{1.2}
\centering
\caption {Summary of GPU Memory Microbenchmark Studies}
\label{tab:literatureReview}
\begin{tabular}
 {|c|c|m{0.68in}|m{1.35in}|}  \hline
  Reference & Year & Device & Scope \\ \hline
  \cite{volkov2008benchmarking} & 2008 & 8800GTX & Architectures and latencies \\
  \hline
  \cite{papadopoulou2009micro,wong2010demystifying} & 2009 & GTX280 & Architectures and latencies \\
  \hline
  \cite{zhang2011quantitative} &2011 &GTX285 &Throughput \\
  \hline
  \cite{baghsorkhi2012efficient} &2012 &Tesla\textsuperscript{TM}C2050&Architectures and latencies \\
  \hline
  \cite{meltzer2013micro} & 2013 & Tesla\textsuperscript{TM}C2070 &Architectures and latencies\\ \hline
  \cite{mei2014benchmarking} & 2014 &GTX560Ti GTX780 &Architectures and latencies \\ \hline
\end{tabular}
\end{table}

We summarize the related GPU microbenchmark work in Table \ref{tab:literatureReview}. Most of the work on memory structure and access latency is based on the P-chase microbenchmark, which was first introduced in \cite{saavedra1992cpu, saavedra1995measuring} (referred to as Saavedra1992 hereafter). Saavedra1992 was originally designed for CPU hardware and was quite successful for various CPU platforms \cite{mcvoy1996lmbench}. Duchateau et al. developed P-ray, a multi-threaded version of P-chase to explore multi-core CPU cache architectures \cite{duchateau2008p-ray}. They exploited false sharing to quickly determine the cache coherence protocol block size. As the multi-threading feature on GPU is different from that of CPU, we cannot apply their method on GPU directly. Volkov and Demmel used P-chase for a very early GPU device, Nvidia 8800GTX, with a relatively simple memory hierarchy \cite{volkov2008benchmarking}. Papadopoulou et al. applied Saavedra1992 to explore the global memory TLB structure \cite{papadopoulou2009micro}. They also proposed a novel footprint experiment (referred to as Wong2010 hereafter) to investigate the other GPU caches \cite{wong2010demystifying}. Baghsorkhi et al. applied Wong2010 to benchmark a Fermi GPU and disclosed its L1/L2 data cache structure \cite{baghsorkhi2012efficient}. Different from previous studies, we proposed a novel fine-grained P-chase and disclosed some unique characteristics that both Saavedra1992 and Wong2010 neglected \cite{mei2014benchmarking}. Our target hardware includes the caches of three recent generations of GPUs.

Meltzer et al. used both Saavedra1992 and Wong2010 to study the  L1/L2 data cache of Fermi architecture \cite{meltzer2013micro}. They found that the L1 data cache does not use the least recently used (LRU) replacement policy, which is one of the basic assumptions of the traditional P-chase \cite{wong2010demystifying}. They also found that the L2 cache associativity is not an integer. Our experimental results coincide with theirs. Moreover, our fine-grained P-chase microbenchmarks allow us to obtain the L1 cache replacement policy.

Zhang and Owens quantitatively benchmarked the global/shared memory throughput from the bandwidth perspective \cite{zhang2011quantitative}. Our work also includes a throughput study, but we are more interested to study the major factors that affect the effective memory throughput. Moreover, we include the study of memory access latencies which are also important factors for performance optimization.



\section{Overview of GPU Memory Hierarchy}

\begin{figure}
\centering
\includegraphics[width=0.47\textwidth]{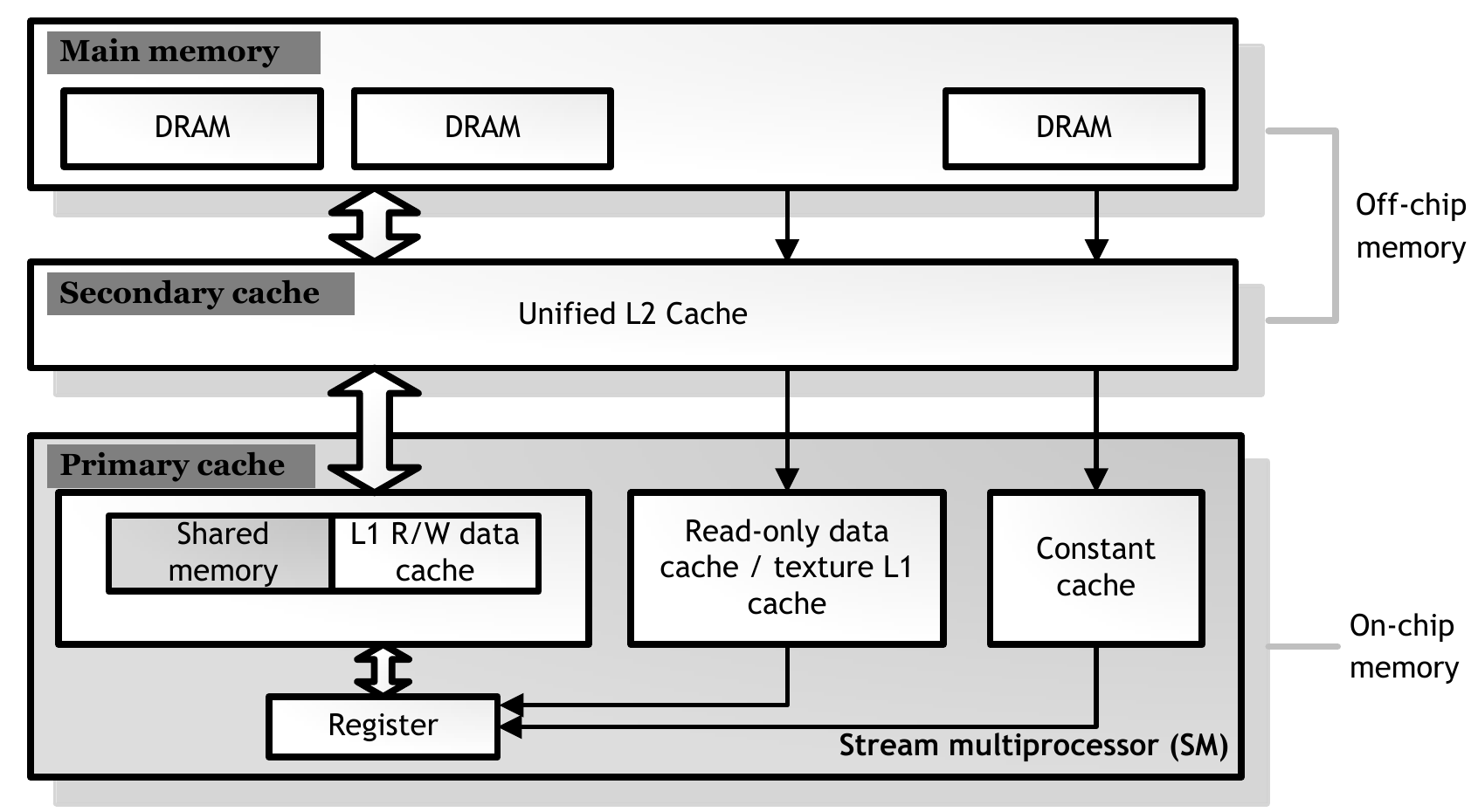}
\caption{Memory hierarchy of the GeForce GTX780 (Kepler).}
\label{fig:KeplerBlock}
\end{figure}

Following the terminologies of CUDA, there are six types of GPU memory space: register, constant memory, shared memory, texture memory, local memory, and global memory. Their properties are elaborated in \cite{cudacprogrammingguide,cudabestguide}. In this study, we limit our scope to the three common types: global, shared, and texture memory. Specifically, we focus on the mechanism of different memory caches, the throughput and latency of global/shared memory, and the effect of bank conflicts on shared memory access latency.

\begin{table}
\renewcommand{\arraystretch}{1.1}
\centering
\caption {Features of Common GPU Memory}
\label{tab:memFeatures}
\begin{tabular}
 {|c|c|c|l|}  \hline  
  Memory & Type & Cached & Scope \\ \hline
  Global & R/W & Yes (CA 2.0 or above) & All threads \\
  \hline
  Shared & R/W & N/A & Thread block\\
  \hline
  Texture &R &Yes & All threads \\
  \hline
\end{tabular}
\end{table}

Table \ref{tab:memFeatures} lists some salient characteristics of the target memory spaces. Unlike the early devices studied in \cite{wong2010demystifying}, in recent GPUs the global memory access has become cached. The cached global/texture memory uses a two-level caching system. The L1 cache is located in each stream multiprocessor (SM), while the L2 cache is off-chip and shared among all SMs. It is unified for instruction, data and page table access. Furthermore, page table is used by GPU to map virtual addresses to physical addresses, and is usually stored in the global memory. The TLB is the cache of the page table. Once a thread cannot find the page entry in the TLB, it would access the global memory to search the page table, which causes significant access latency. Although the global memory and texture memory have similar dataflows, the former is read-and-write (R/W) and the latter is read-only. Both of them are public to all threads in the kernel function. The GPU-specific R/W shared memory is also located in the SMs. On the Fermi and Kepler devices it shares memory space with the L1 data cache, whereas on the Maxwell devices it has a dedicated space. In CUDA, the shared memory is declared and accessed inside a cooperative thread array (CTA, a.k.a. thread block), which is a programmer-assigned set of threads executed concurrently. Fig. \ref{fig:KeplerBlock} shows the block diagram of the memory hierarchy of a Kepler device, GeForce GTX780. The arrows indicate the dataflow. The architecture of the L1 cache in the Maxwell device is slightly different from that shown in Fig. \ref{fig:KeplerBlock} due to the separate shared memory space.

\begin{table*}
\centering
\renewcommand{\arraystretch}{1.1}
\caption{Comparison of the Memory Properties of the Tesla, Fermi, Kepler and Maxwell Devices}
\label{tab:comparison}

\begin{tabular}
{|l|m{1in}<{\centering}|m{1in}<{\centering}|m{1in}<{\centering}|m{1in}<{\centering}|}
\hline
\multirow{2}{*}{Device} & \tc{Tesla} & \tc{Fermi }& Kepler & Maxwell \\
& GTX280 & GTX560Ti & GTX780 & GTX980 \\
\hline
Compute capability& 1.3 & 2.1 &  3.5 & 5.2\\ \hline
SMs * cores per SM& 30 * 8 & 8 * 48 & 12 * 192 & 16 * 128 \\
\hline \hline

\multicolumn{5}{|c|}{Global memory} \\
\hline
Cache mechanism & N/A & L1 and L2 & L2, or read-only & L2, or unified L1\\
\hline
\multirow{2}{*}{Cache size} &\multirow{2}{*} {N/A} & L1: 16/48 KB& Read-only: 12 KB&Unified L1: 24 KB\\
&&L2: 512 KB &L2: 1.5 MB&L2: 2 MB\\
\hline
Total size & 1024 MB & 1024 MB &3072 MB &4096 MB \\
\hline \hline

\multicolumn{5}{|c|}{Shared memory} \\
\hline
Size per SM & 16 KB & 48/16 KB & 48/32/16 KB & 96 KB \\
\hline
Maximum size per CTA & 16 KB &\multicolumn{3}{c|}{48 KB}\\
\hline
Bank No. & 16 &\multicolumn{3}{c|}{32}\\
\hline
Bank width & \multicolumn{2}{c|}{4 B} & 8 B &4 B\\
\hline \hline

\multicolumn{5}{|c|}{Texture memory} \\
\hline
Texture units & per-TPC & \multicolumn{3}{c|}{per-SM}\\
\hline
L1 cache size & 6-8 KB & 12 KB & 12 KB & 24 KB \\
\hline
\end{tabular}

\end{table*}

In Table \ref{tab:comparison}, we compare the memory characteristics of the old Tesla GPU discussed in \cite{papadopoulou2009micro,wong2010demystifying} and our three target GPU platforms. The compute capability is used by NVIDIA to distinguish the generations. Table \ref{tab:comparison} shows that the most distinctive difference lies in the global memory. On the Tesla device, the global memory access is not cached, whereas on the Fermi device it is cached in both the L1 and the L2 data cache. The Kepler device has an L1 data cache, but it is designed for local rather than global memory access. In addition to the L2 data cache, global memory data that is read-only for the entire lifetime of a kernel can be cached in the read-only data cache with a compute capability of 3.5 or above. On the Maxwell device, the L1 data cache, texture on-chip cache and read-only data cache are combined in one physical space. Note that the L1 data cache of the Fermi and the read-only data cache of the Maxwell can be turned on or off. It is also notable that modern GPUs have larger shared memory spaces and more shared memory banks. On the Tesla device, the shared memory size of each SM is fixed at 16 KB. On the Fermi and Kepler devices, the shared memory and L1 data cache share 64 KB of memory space. On the Maxwell device, the shared memory is independent and has 96 KB. The maximum volume of shared memory that can be assigned to each CTA has been increased from 16 KB on the Tesla device to 48 KB on the later devices. The texture memory is cached on every generation of GPUs. The Tesla texture units are shared by three SMs (i.e., thread processing cluster). However, texture units on later devices are per-SM. The texture L2 cache shares space with the L2 data cache. The size of the texture L1 cache depends on the generation of the GPU hardware.
%


\section{Cache Structures}

The greatest difference between recent GPUs and the old Tesla GPUs lies in their cache systems. In this section, we first present a novel fine-grained P-chase method, and then explore two kinds of cache: the data cache and the TLB. We focus on the architectures of the Fermi/Maxwell L1 data cache, Fermi/Kepler/Maxwell texture memory L1 cache, read-only data cache, L2 cache and TLBs.

\subsection{Why Not Typical P-chase?}

\begin{figure}
  \centering
  \includegraphics[width=0.48\textwidth]{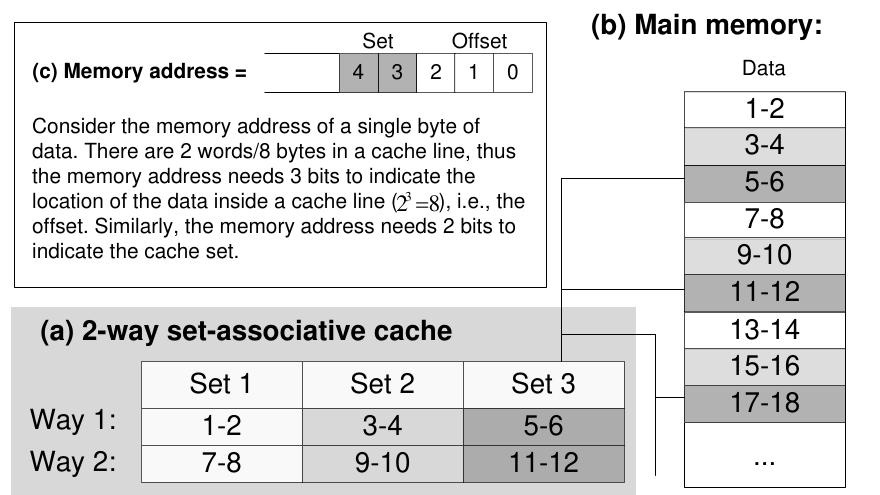}
  \caption{An example of a 12-word 2-way set-associative cache. Assume each word has 4 bytes, each cache line can store 2 words ($b=2$), and the data array is sequentially accessed. The cache lines are grouped into 3 separate cache sets ($T=3$), each of which has 2 cache lines (i.e., Way 1 and Way 2), and we say its \emph{cache associativity} is 2 ($a=2$). }
  \label{fig:covCache}
\end{figure}

By exploiting the principle of locality, cache memory is used to back up a piece of main memory for faster data access and plays a major role in modern computer architectures. Most existing GPU microbenchmark studies on cache architecture assume a classical set-associative cache model with the least recently used (LRU) replacement policy, the same as that of a conventional CPU cache \cite{saavedra1992cpu, saavedra1995measuring}. The cache size ($C$) is much smaller than main memory size. Data is loaded from main memory to cache with the basic unit of a cache line. The number of words in a cache line is referred to as the line size ($b$). For the classical \emph{LRU set-associative cache}, the cache memory is divided into $T$ cache sets, each of which consists of $a$ cache lines. Fig. \ref{fig:covCache} shows an example of a 12-word set-associative cache and its memory mapping. There are three essential assumptions for this kind of cache model:

\begin{cor}
\label{def:cond1}
All cache sets have the same size, and the cache parameters satisfy $T*a*b=C$. If any three of the four parameters are known, the remaining one can be found.
\end{cor}

\begin{cor}
\label{def:cond2}
In the memory address, the bits that identify the cache set are immediately followed by the bits that identify the offset (the intra-cache line location of data).
\end{cor}

\begin{cor}
\label{def:cond3}
The cache replacement policy is LRU.
\end{cor}

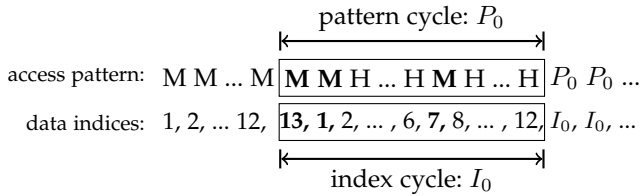
\begin{figure}
\centering
\begin{tikzpicture}[scale=0.49]

\node[left] at (0,0) {\footnotesize data indices:} ;
\node[left] at (0,1.2) {\footnotesize access pattern:};

\node[right] at (0,0) {\small 1, 2, ... 12,};
\node[right] at (0,1.2) {M M ... M};

\draw (3.4,-0.5) rectangle (10.6,0.5);
    \node at (7,0) {\small \textbf{13, 1,} 2, ... , 6, \textbf{7,}  8,  ... , 12,};
\draw (3.4,0.7) rectangle (10.6,1.7);
    \node at (7,1.2) {\textbf{M M} H ... H \textbf{M} H ... H};

\draw [thick, black, |<->|] (3.4, 2.2)--node[above] {pattern cycle:  $P_0$ } (10.6,2.2);
\draw [thick, black, |<->|] (3.4, -1)--node[below] {index cycle: $I_0$ } (10.6,-1);
\node[right] at (10.5,0) {\small $I_0$, $I_0$, ...};
\node[right] at (10.5,1.2) {$P_0$ $P_0$ ...};
\end{tikzpicture}
\vspace{-1em}
\caption{Periodic memory access pattern of a classical LRU set-associative cache. M means cache miss and H means cache hit.}
\label{fig:PeriodicOutput}
\end{figure}

Assumption \ref{def:cond1} implies that all cache sets have the same number of cache lines. Assumption \ref{def:cond2} indicates that the data mapping from the main memory to the cache follows a predictable, regular pattern. For instance, in Fig. \ref{fig:covCache}, two out of every six consecutive words are mapped to one cache set, and they may appear in either of the two cache lines in the set. Assumption \ref{def:cond3} implies that if we perform sequential loading of a piece of data, the memory access is periodic. Taking the cache model in Fig. \ref{fig:covCache} as an example, we initialize an array with 13 words and read it one word by one word. Fig. \ref{fig:PeriodicOutput} shows the full memory access process and its access pattern (a cache miss or a cache hit generated by visiting one array element). As the array size is one word larger than the cache size, the cache miss occurs. With the exception of the first 12 data accesses, which are cold cache misses, those data accesses to the 1st, 7th and 13th array elements are cache misses while the rest are cache hits. The 13-25th memory accesses form a pattern $P_0$, which recurs until the end of the data loading process. The period of this memory access pattern is 13, which equals the array length.

\begin{table}
\centering
\caption{Notations for Cache and P-chase Parameters}
\label{tab:notation}
    \begin{tabular}{|c|c ||c|c |}
    \hline
    Notation & Description & Notation & Description\\
    \hline
    $C$ & cache size & $N$ & array size \\
    $b$ & cache line size & $s$ & stride size \\
    $a$ & cache associativity & $k$ & iterations \\
    $T$ & number of cache sets & $r$ & cache miss rate \\
    \hline
    \end{tabular}
\end{table}

\renewcommand{\thelstlisting}{\arabic{lstlisting}}
\begin{lstlisting}[caption={P-chase: array initialization}, label={list:P-chase_initial}]
for(i=0;i<array_size;i++){
   A[i]=(i+stride)%array_size;
}
\end{lstlisting}

\renewcommand{\thelstlisting}{\arabic{lstlisting}}
\begin{lstlisting}[ caption={P-chase: kernel function}, label={list:P-chase_chase}]
start_time = clock();
for(it=0;it<iterations;it++){
    j=A[j];
}
end_time=clock();
//calculate average memory latency
tvalue=(end_time-start_time)/iterations;
\end{lstlisting}

The P-chase microbenchmark is a successful method for obtaining cache parameters \cite{saavedra1992cpu,saavedra1995measuring,mcvoy1996lmbench,volkov2008benchmarking,
papadopoulou2009micro,wong2010demystifying,baghsorkhi2012efficient,meltzer2013micro}. The core idea of P-chase is to traverse an array whose elements are initialized as the indices for the next memory access. The distance between two consecutively accessed array elements is called \emph{stride} and is usually fixed in an experiment. The memory access latency is highly dependent on the \emph{stride} due to the cache effect. By measuring \textbf{the average memory access latency} of a great number of memory accesses, the cache parameters can be deduced from the array size and the stride size. Listing \ref{list:P-chase_initial} and Listing \ref{list:P-chase_chase} give the array initialization and the kernel function of P-chase. In Listing \ref{list:P-chase_chase}, \emph{j}=\emph{A}[\emph{j}] is repeatedly executed over $iterations$ of times, so that the array $A$ is sequentially traversed with a fixed \emph{stride}. Before the timing, we load the array elements for a number of times to eliminate the cold instruction cache misses. The average memory access latency, $tvalue$, is calculated by dividing the total clock cycles by \emph{iterations}. We denote the array size, stride size, and $iterations$ by $N$, $s$ and $k$, respectively. We summarize the notations in Table \ref{tab:notation}.

Based on Assumptions 1-3, the output of P-chase, i.e., the average memory access latency, $t_{avg}$, satisfies
\[t_{avg}=t_0*(1-r)+(t_0+t_m)*r=t_0+t_m*r\]
where $r$ denotes the cache miss rate, $t_0$ denotes the cache access latency and $t_m$ denotes the cache miss penalty. Because $t_0$ and $t_m$ are hardware-dependent constants, the typical P-chase method actually relies on the cache miss rate, $r$.

It has been believed that the cache parameters can be deduced from the $tvalue$-$s$ graph (Saavedra1992) or the $tvalue$-$N$ graph (Wong2010). As mentioned, under Assumptions 1-3, the memory access, or the cache miss patterns are periodic. Moreover, both Saavedra1992 and Wong2010 suggest that not only the cache miss patterns are predictable, but also the possible values of $r$ are predictable.

In particular, Saavedra1992 suggests to run the experiments for multiple times, each with a different stride. Both array size $N$ and stride size $s$ are usually set to be power-of-two. If $N$ is much larger than the cache size $C$, and $s$ is smaller than cache line size $b$, there is a cache miss when loading the data mapped to the beginning address of a cache line, i.e., the cache miss rate is $s/b$. If $s\ge b$ but not exceeding $N/a$, every data loading is a cache miss. When $s$ continues growing, the loaded data can fit into the cache so that there is no cache miss. To summarize, the cache miss rate satisfies Eq. (\ref{eq_1}) for all $(N, s)$ pairs.
\begin{equation}
r \in \{0, s/b, 1\}, N>>C \label{eq_1}
\end{equation}
Wong2010 suggests visiting arrays of various sizes with a fixed stride, which is chosen carefully and should be around cache line size. If we choose $s=b$, then every time we increase array size by $b$, there are much more cache misses. The cache miss rate satisfies Eq. (\ref{eq_2}) for all $(N, s)$ pairs.
\begin{equation}
r \in \{0, \frac{1}{T}, ..., \frac{k}{T},..., 1\}, N \in [C, C+T*b], s=b \label{eq_2}
\end{equation}
~
Fig. \ref{fig:P-chase_sample_92} and Fig. \ref{fig:P-chase_sample_wong} show the experimental results when we apply Saavedra1992 and Wong2010 on the texture L1 cache on GTX780. Surprisingly, we obtain different results from the two methods. In Fig. \ref{fig:P-chase_sample_92}, the $N$=12KB line suggests that $C=12$ KB. The $N$=48KB line at $log_2(s)=5$ suggests $b=32$ bytes, and at $log_2(s)=11$ suggests $a=N/s=24$ so that $T=C/(ab)=16$. In Fig. \ref{fig:P-chase_sample_wong}, there are 4 plateaus between the minimum and maximum memory latency, which indicates there are 4 cache ways in a cache set. The cache line size equals the width of every plateau. Overall, it suggests that $C=12$ KB, $b=128$ bytes, $T=4$, and $a=C/(bT)=24$. Here we face a contradiction: Fig. \ref{fig:P-chase_sample_92} and Fig. \ref{fig:P-chase_sample_wong} are based on the same hardware, yet they lead to different cache parameters. This motivates us to seek the underlying causes.

\begin{figure}
\centering
\begin{minipage}[b]{0.47\linewidth}
\tikzstyle{every pin}=[
	font=\scriptsize]
\begin{tikzpicture}[scale=0.7]
    \begin{axis}[small,
            width=1.4\textwidth,
            xlabel={$log_2 s$ },
            ylabel={Latency (clock cycles)},
            ymax=250 , ymin = 0,
            xmajorgrids=true,
            ymajorgrids=true,
            legend pos=south east,font=\scriptsize,
            xmin=1,xmax=15,xtick={1,2,3,...,15}]
        \addplot
            table[x=stride,y=latency] {./data/texture_12KB.txt};
        \addplot
            table[x=stride,y=latency] {./data/texture_48KB.txt};
        \addplot[dashed,thick,|<->|]coordinates {(3,0) (3,108)}
	          node[left] at (axis cs:3,50) {$t_0$};
        \node [coordinate,pin=330:{cache line size}] at (axis cs: 5,228) {};
        \node [coordinate,pin=120:{cache associativity}] at (axis cs: 11,108) {};
    \legend{N=12KB, N=48KB}
    \end{axis}
\end{tikzpicture}
\vspace{-1em}
\caption{$tvalue$-$s$ of the Kepler texture L1 data cache.}
\label{fig:P-chase_sample_92}
\end{minipage}
~
\begin{minipage}[b]{0.48\linewidth}
\tikzstyle{every pin}=[
	font=\scriptsize]
\begin{tikzpicture}[scale=0.7]
    \begin{axis}[small,
            width=1.4\textwidth,
            xlabel={Array size, $N$ (KB)},
            ylabel={Latency (clock cycles)},
            ymax=240 , ymin =100,
            xmajorgrids=true,
            ymajorgrids=true,
            xmin=11.95,xmax=12.6,xtick={12,12.125,12.25,12.375,12.5}]
        \addplot
            table[x=array_size,y=latency] {./data/texture_footprint.txt};

        \addplot[dashed,thick,|<->|]coordinates {(12,220) (12.5,220)}
	          node[above] at (axis cs:12.2,220) {\scriptsize{way size}};
        \addplot[dashed,thick,|<->|]coordinates {(12.375,180) (12.5,180)}
	          node[below] at (axis cs:12.43,180) {\scriptsize {cache line size}};
        \node [coordinate,pin=30:{cache size}] at (axis cs: 12,108) {};
    \end{axis}
\end{tikzpicture}
  \caption{$tvalue$-$N$ of the Kepler texture L1 data cache (8-byte stride).}
  \label{fig:P-chase_sample_wong}
\end{minipage}
\end{figure}

Both Saavedra1992 and Wong2010 methods are based on Assumptions 1-3 so that the cache miss rates satisfy Eqs. (\ref{eq_1}) and (\ref{eq_2}). However, our experimental results reveal that Assumptions 1-3 seldom hold for different types of GPU cache, consequently Eqs. (\ref{eq_1}) and (\ref{eq_2}) are ineffective. Thus, the typical P-chase results become inappropriate to expose the GPU cache structure. For example, if Assumptions 1 and 2 hold but Assumption 3 does not, and the cache replacement policy is random, then the measured $t_{avg}$ can vary even for a given $(N, s)$ pair. The value of $r$ also varies and may not belong to those listed in (1) or (2). Hence, $t_{avg}$ alone fails to serve as an indicator of GPU cache architecture.

Motivated by the above observation, we designed a microbenchmark that utilizes GPU shared memory to display the latency of every single memory access. We refer to it as fine-grained P-chase microbenchmark because it provides the most detailed information on the data access process.

\subsection{Our Methodology: Fine-grained P-Chase}

\renewcommand{\thelstlisting}{\arabic{lstlisting}}
\begin{lstlisting}[caption={Fine-grained P-chase kernel (single thread, single CTA)}, label={list_kernelcode}]
__global__ void KernelFunction(...){
  //declare shared memory space
  __shared__ unsigned int s_tvalue[];
  __shared__ unsigned int s_index[];
  preheat the data;
  for(it=0;it<iterations;it++) {
    start_time=clock();
    j=my_array[j];
    //store the array index	
    s_index[it]=j;
    end_time=clock();
    //store the access latency
    s_tvalue[it]=end_time-start_time;
  }
}
\end{lstlisting}

The core idea of our fine-grained P-chase is to record and analyze every single data access latency in a kernel with a single thread and single CTA. Such method is difficult to be used for CPU cache because of the challenge of recording every data access latency without interfering the normal data access. However, we can exploit GPU shared memory to store a sequence of data access latencies, based on which we can deduce the cache structure and parameters. The shared memory access is prompt and does not affect the data cache. Listing \ref{list_kernelcode} gives the kernel code of our single-thread fine-grained P-chase. Notice that before the measurement, we need to visit the data in an initial iteration, aiming to load the data into L2 cache. Doing so can avoid the cold instruction cache miss and the interference from possible hardware pre-fetching. The core statement in line 8, \emph{j = my\_array}[\emph{j}], is the same as in the conventional P-chase. The difference lies in the location of the timing function. We put the timing statements inside a long loop, as shown in lines 7 and 11. The \emph{clock}() function provided by CUDA is implemented by reading a special register, the value of which is incremented every clock cycle. We measure the overhead of \emph{clock}() as the difference between two consecutive \emph{clock}() calls in a single kernel thread. Based on our experimental results, the overhead of \emph{clock}() is 14, 16, and 6 cycles on Fermi, Kepler, and Maxwell platforms, respectively.

Although the idea of fine-grained P-chase is simple, we need to address the following major challenge: due to instruction-level parallelism (ILP), function \emph{clock}() may overlap with its previous instruction and even return before the previous instruction finishes. E.g., if we put the second \emph{clock}() (line 11) immediately after statement \emph{j = my\_array}[\emph{j}] (line 8), it may lead to incorrect memory latency measurements because the second \emph{clock}() could return before line 8 finishes. We overcome this problem by introducing a new statement, \emph{s\_index}[\emph{it}] = \emph{j} (line 10), that has data dependency on line 8, to ensure that the memory access completes when line 11 is issued. We use a separate program to measure the overhead of the code segment of lines 10-11, which is 20, 32, 16 cycles on Fermi, Kepler, and Maxwell, respectively. We can then deduce the latency of line 8 alone.

Our fine-grained P-chase microbenchmark outputs two arrays, \emph{s\_tvalue}[] and \emph{s\_index}[], the lengths of which are equal to the value of $iterations$. The former contains the data access latencies and the latter contains the accessed data indices. With these two arrays, we can reproduce the entire memory loading process and obtain all of the data access latencies rather than the average.


\begin{figure*}
  \centering
  \includegraphics[width=0.7\textwidth]{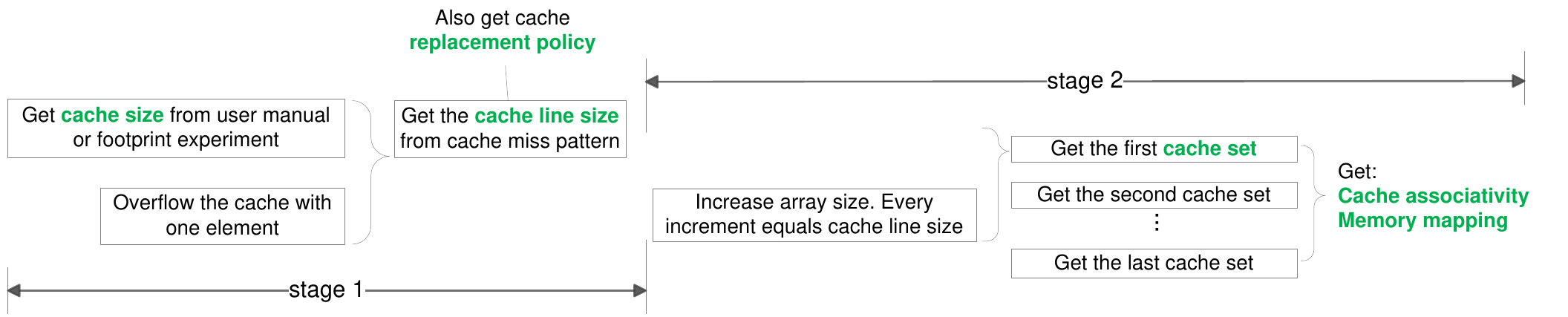}
  \caption{Flowchart of applying fine-grained P-chase.}\label{fig:flowchart}
\end{figure*}

We work out a procedure to find the cache parameters using our fine-grained P-chase microbenchmark with different $(N, s)$ configurations. Fig. \ref{fig:flowchart} shows the flowchart of our two-stage procedure. We could use brute-force $N$ testing to get the cache size. Then in the first stage, we overflow the cache with one element, getting the cache line size. We can also find whether the cache replacement policy is LRU or not in this stage. In the second stage, we gradually overflow the cache with the granularity of a cache line, until all the data accesses become cache miss. We can deduce the cache associativity and the memory addressing from the second stage. We further elaborate our method as follows. Notice that the basic unit of $(N, s)$ is the length of an array element.

\begin{enumerate}
\item Determine cache size $C$. We set $s$ to 1. We then initialize $N$ with a small value and increase it gradually until the first cache miss appears. $C$ equals the maximum $N$ where all memory accesses are cache hits.
\item Determine cache line size $b$. We set $s$ to 1. We begin with $N=C+1$ and increase $N$ gradually again. When $N<C+b+1$, the numbers of cache misses are close. When $N$ is increased to $C+b+1$, there is a sudden increase on the number of cache misses, despite that we only increase $N$ by 1. Accordingly we can find $b$. Based on the memory access patterns, we can also have a general idea on the cache replacement policy.
\item Determine number of cache sets $T$. We set $s$ to $b$. We then start with $N=C$ and increase $N$ at the granularity of $b$. Every increment causes cache misses of a new cache set. When $N>C+(T-1)b$, all cache sets are missed. We can then deduce $T$ from cache miss patterns accordingly.
\item Determine cache replacement policy. As mentioned before, if the cache replacement policy is LRU, then the memory access process should be periodic and all the cache ways in the cache set are missed. If memory access process is aperiodic, then the replacement policy cannot be LRU. Under this circumstance, we set $N=C+b, s=b$ with a considerable large $k$ ($k>>N/s$) so that we can traverse the array multiple times. All cache misses are from one cache set. Every cache miss is caused by its former cache replacement because we overflow the cache by only one cache line. We have the accessed data indices thus we can reproduce the full memory access process and find how the cache lines are updated.
\end{enumerate}

Applying the above method, we sketch the structures of texture L1 cache, read-only data cache, L1/L2 TLBs, and on-chip L1 data cache in the following sections. We also present some preliminary results of the off-chip L2 data cache.



\subsection{Texture L1 Cache and Read-only Data Cache}

\begin{figure*}
\centering
  \includegraphics[width=0.68\textwidth]{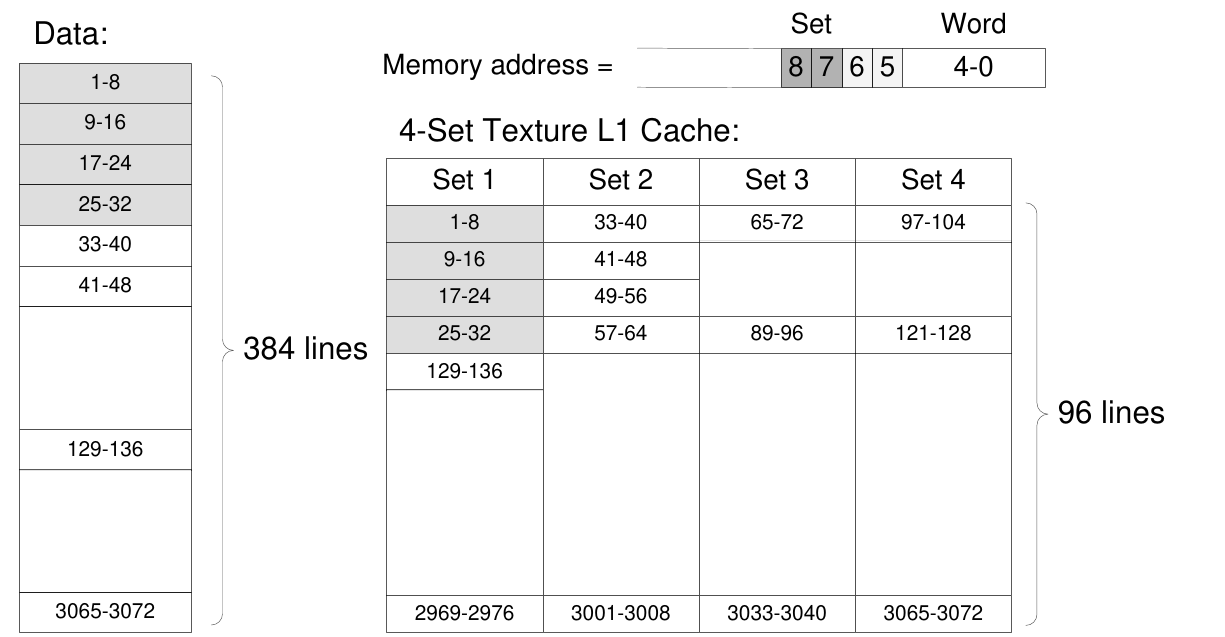}
\caption{The texture L1 cache structure of the Fermi and Kepler device and the memory address.}
  \label{fig:textureL1cacheMapping}
\end{figure*}

\begin{figure*}
\centering
\begin{minipage}[b]{0.32\textwidth}
\begin{tikzpicture}[scale=0.86]
        \begin{axis}[
            cycle list name=black white,
            width=1.15\textwidth,
            height=0.8\textwidth,
            xlabel={Array size (MB)},
            ylabel={Cache miss rate (\%)},
            ymin=-5, ymax=105 ,
            xmajorgrids=true,ymajorgrids=true,
            xtick={130,132,134,136,138,140,142,144},
            legend pos= south east,font=\footnotesize,
        ]
            \addplot
            coordinates{
            (128, 0)
            (130, 0)
            (132, 1700/65)
            (134, 2500/65)
            (136, 3300/65)
            (138, 4100/65)
            (140, 4900/65)
            (142, 5700/65)
            (144, 100)
            (146, 100)
            };
            \addplot
            coordinates{
            (128, 0)
            (130, 0)
            (132, 100/7)
            (134, 200/7)
            (136, 300/7)
            (138, 400/7)
            (140, 500/7)
            (142, 600/7)
            (144, 100)
            (146, 100)
            };
        \legend{measured,expected}
        \end{axis}
        \end{tikzpicture}
        \vspace{-1em}
\caption{Miss rate of L2 TLB (2 MB stride).}
\label{fig:tlb_compare}
\end{minipage}
~
\begin{minipage}[b]{0.31\textwidth}
\centering
\includegraphics[width=0.98\textwidth]{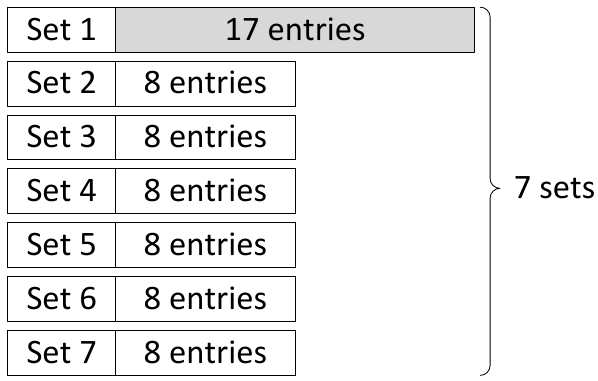}
\vspace{-1em}
\caption{L2 TLB structure.}\label{fig:L2TLB_structure}
\end{minipage}
~
\begin{minipage}[b]{0.33\textwidth}
\begin{tikzpicture}[scale=0.36]
    \draw[fill=gray!40] (3,0) rectangle (6,8);
    \foreach \x[evaluate={\i=int(1+\x/3)}] in {0,3,6,9} {
        \draw[very thick] (\x,0) rectangle (\x+3, 8);
        \node at (\x+1.5, -1) {Way \i};
        }
    \draw (0,0) grid[xstep=3, ystep=1] (12,8);
    \foreach \y in {0,1,2,...,7}
    \foreach \x[evaluate={\i =int(32*(\x/3)+1+4*\y); \j= int(\i+3)}] in {0,3,6,9}
        \node[right] at (\x,\y+0.5) {\scriptsize \i-\j};
    \draw[thick, |<->|] (12.5,0)--node[below,rotate=90]{32 cache sets} (12.5,8);
    \draw[thick, |<->|] (-0.5,7)-- (-0.5,8);
    \node[rotate=90] at (-1.2, 5.5){A major set};
\end{tikzpicture}
\vspace{-1em}
\caption{L1 data cache structure (16 KB).}
\label{fig:fermi_L1replacement}
\end{minipage}

\end{figure*}

We apply our P-chase microbenchmark on the texture L1 cache with the two-stage methodology. We bind an unsigned integer array to the linear texture, and fetch it with \emph{tex1Dfetch}(). In the first stage, we find out the cache size $C$, which is 12 KB; and then set $s$ to 1 element (i.e., 4 bytes) and overflow the cache gradually to get the cache line size, which is 32 bytes. In the second stage, we increase $N$ from 12 KB to 12.5 KB with $s=32$ bytes. Our results suggest a 12 KB set-associative cache with a special memory address format, as shown in Fig. \ref{fig:textureL1cacheMapping}, on Fermi and Kepler devices, and a 24 KB cache with similar organization on Maxwell device.

On the Fermi and Kepler GPUs, the 12 KB texture L1 cache is divided to 4 cache sets and can store up to 384 cache lines. Each cache set contains 96 cache lines and each cache line contains 8 words (i.e., 32 bytes). Each consecutive 32 words (i.e., 128 bytes) is mapped onto 4 successive cache sets. In particular, the 7-8th bits of the memory address determine the corresponding cache set, whereas the 5-6th bits do so in the traditional set-associative cache design. This mapping is optimized for 2D spatial locality in graphic processing \cite{cudabestguide,hakura1997design}. To take advantage of this mapping, in generalized applications, threads within a warp need to visit adjacent memory addresses, otherwise there would be more cache misses. The Maxwell texture L1 cache has a similar structure except it contains 768 cache lines.


Devices with a compute capability of 3.5 or above have an on-chip per-SM read-only data cache, which is an improvement on the texture memory cache \cite{keplerwhitepaper}. The read-only data cache is loaded by calling $\_\_ldg(const\ \_\_restricted\_\_ * address)$. On our GTX780, we find a 12 KB read-only data cache, the same as the texture L1 cache. We overflow the read-only data cache with a single 4-byte element and find that the cache line size is 32 bytes and the replacement policy is LRU. We then examine it with $s=32$ bytes and $N$ varying from 12 KB to 60 KB. When the array is larger than 12.5 KB, each data access results in a cache miss. We infer that the read-only cache structure is the same as the texture L1 cache: 4 cache sets, with a 32-byte cache line and 96 lines in each set. Similarly, 128 successive bytes are mapped onto the same set, but the data mapping is not bits-defined. On the GTX980, the structure of the read-only data cache is also the same as that of the texture L1 cache except for the rather random data mapping.


\subsection{Translation Look-Aside Buffer}

Previous studies show that the GTX280 has two levels of TLB to support GPU virtual memory addressing on the GTX280 \cite{papadopoulou2009micro,wong2010demystifying}, where the L1 TLB is 16-way fully associative and the L2 TLB is 8-way set-associative. We apply our fine-grained P-chase method to investigate the TLB of three recent GPU architectures and find them have the same 16-way fully associative L1 TLB, and the page size is 2 MB. We plot the cache miss rate of the L2 TLB in Fig. \ref{fig:tlb_compare} based on our microbenchmark results. The traditional LRU cache with equal sets triggers the same number of cache misses each time, thus the expected cache miss rate increases linearly. In contrast, our measured miss rate increases piecewise linearly. When $N$ equals 132 MB, we observe 17 missed entries; varying $N$ from 134 MB to 144 MB with $s=2$ MB causes 8 more missed entries each time. Considering that cache misses are triggered set by set, the only explanation for the piecewise linear increase is that the first cache set has more cache ways than others. In addition, we deduce that the replacement policy is LRU, as the number of cache ways is equal to the number of missed cache entries. This gives us the conjectured L2 TLB structure as shown in Fig. \ref{fig:L2TLB_structure}: 1 large set with 17 entries and 6 small sets with 8 entries each.

\begin{table*}
\renewcommand{\arraystretch}{1.1}
\centering
\caption{Parameters of Common GPU Caches}
\begin{tabular}{|c|m{0.8in}<{\centering}|m{0.8in}<{\centering}|m{0.8in}<{\centering}|m{1.1 in}<{\centering}|m{1.1 in}<{\centering}|}
    \hline
    Parameters & Default Fermi L1 data cache & Fermi/ Kepler/ Maxwell L1 TLB & Fermi/ Kepler/ Maxwell L2 TLB & Fermi/ Kepler texture L1 cache/ Kepler read-only data cache& Maxwell L1 data/ texture L1 cache/ read-only data cache \\ \hline
    $C$ & 16 KB  & 32 MB & 130 MB & 12 KB & 24 KB\\ \hline
    $b$ & 128 byte & 2 MB & 2 MB & 32 byte & 32 byte \\ \hline
    $T$ & 32 & 1 & 7 & 4  &4 \\ \hline
    LRU & no & no & yes & yes &yes\\ \hline
\end{tabular}
\label{tab:parameters}

\end{table*}

\subsection{L1 Data Cache}
On the Fermi and Kepler devices, the L1 data cache and shared memory are physically implemented together. On the Maxwell devices, the L1 data cache is unified with the texture cache.

The Fermi L1 data cache can be either 16 KB or 48 KB. We only report the 16 KB case here for brevity. We vary the array size from 15 KB to 24 KB with $s=4$ bytes or $s=128$ bytes, and observe the memory access patterns. Fig. \ref{fig:fermi_L1replacement} gives the Fermi 16 KB L1 cache structure based on our experimental results. The 16 KB L1 cache has 128 cache lines mapped onto four cache ways. For each cache way, 32 cache sets are divided into 8 major sets. Each major set contains 16 cache lines. The data mapping is also unconventional. The 12-13th bits in the memory address define the cache way, the 9-11th bits define the major set, and the 0-6th bits define the memory offset inside the cache line.

One distinctive feature of the Fermi L1 cache is that its replacement policy is not LRU, as pointed out by Meltzer et al. in \cite{meltzer2013micro}. In our experimental results, the memory access process does not reveal periodicity. We demonstrate part of the memory access process with $N=16.125$ KB (i.e., 129 data lines), $s=128$ bytes in Fig. \ref{fig:ProcessL1}. Because we overflow the cache with only one line, all cache misses are from a single cache set. In our experiment, cache misses occur when accessing data line 3, 35, 68, 100 and 129, which therefore belong to the first cache set. When we read the 129th data line, it sometimes leads to a cache miss and sometimes a cache hit. This cannot happen in the conventional LRU cache model. We find that among the four cache ways, cache way 2 is three times more likely to be replaced than the other three cache ways. It is updated once every two cache misses. The replacement probabilities of the four cache ways are $\frac{1}{6}, \frac{1}{2}, \frac{1}{6}$ and $\frac{1}{6}$, respectively.

For sequential data loading in our experiment, this non-LRU cache reduces the number of cache misses compared with the conventional cache; for example, in Fig. \ref{fig:ProcessL1}, the listed memory accesses should all be cache misses if the LRU replacement policy were used.

\begin{figure}
\centering
\begin{tikzpicture}[scale=0.48]
\foreach \x[evaluate={\i=int(\x/2-1)}] in {4,6,8,10}
        \node at (\x+1, 0.5) {\small Way\i};
\draw (4,0) grid [xstep=2, ystep=-1] (12,-1);
\node at (5, -0.5) {\footnotesize 3};
\node at (7, -0.5) {\footnotesize 35};
\node at (9, -0.5) {\footnotesize 68};
\node at (11, -0.5) {\footnotesize 100};
\node at (13, -0.5) {\footnotesize 129};
\node[right] at (-0.5,-0.5) {\small $1^{st}$ cache set: };
\node[right] at (0,-1.5) {read 129:};
\node at (5, -1.5) {\footnotesize 3};
\node[fill=gray!40] at (7, -1.5) {\footnotesize 129};
\node at (9, -1.5) {\footnotesize 68};
\node at (11, -1.5) {\footnotesize 100};
\node at (15, -1.5) {miss};
\node[right] at (0,-2.5) {read 3:};
\node at (5, -2.5) {\footnotesize 3};
\node at (7, -2.5) {\footnotesize 129};
\node at (9, -2.5) {\footnotesize 68};
\node at (11, -2.5) {\footnotesize 100};
\node at (15, -2.5) {hit};
\node[right] at (0,-3.5) {read 35:};
\node at (5, -3.5) {\footnotesize 3};
\node at (7, -3.5) {\footnotesize 129};
\node[fill=gray!40] at (9, -3.5) {\footnotesize 35};
\node at (11, -3.5) {\footnotesize 100};
\node at (15, -3.5) {miss};
\node[right] at (0,-4.5) {read 68:};
\node at (5, -4.5) {\footnotesize 3};
\node[fill=gray!40] at (7, -4.5) {\footnotesize 68};
\node at (9, -4.5) {\footnotesize 35};
\node at (11, -4.5) {\footnotesize 100};
\node at (15, -4.5) {miss};
\node[right] at (0,-5.5) {read 100:};
\node at (5, -5.5) {\footnotesize 3};
\node at (7, -5.5) {\footnotesize 68};
\node at (9, -5.5) {\footnotesize 35};
\node at (11, -5.5) {\footnotesize 100};
\node at (15, -5.5) {hit};

\node[right] at (0,-6.5) {read 129:};
\node[fill=gray!40] at (5, -6.5) {\footnotesize 129};
\node at (7, -6.5) {\footnotesize 68};
\node at (9, -6.5) {\footnotesize 35};
\node at (11, -6.5) {\footnotesize 100};
\node at (15, -6.5) {miss};
\node[right] at (0,-7.5) {read 3:};
\node at (5, -7.5) {\footnotesize 129};
\node[fill=gray!40] at (7, -7.5) {\footnotesize 3};
\node at (9, -7.5) {\footnotesize 35};
\node at (11, -7.5) {\footnotesize 100};
\node at (15, -7.5) {miss};
\node[right] at (0,-8.5) {read 35:};
\node at (5, -8.5) {\footnotesize 129};
\node at (7, -8.5) {\footnotesize 3};
\node at (9, -8.5) {\footnotesize 35};
\node at (11, -8.5) {\footnotesize 100};
\node at (15, -8.5) {hit};
\node[right] at (0,-9.5) {read 68:};
\node at (5, -9.5) {\footnotesize 129};
\node at (7, -9.5) {\footnotesize 3};
\node at (9, -9.5) {\footnotesize 35};
\node[fill=gray!40] at (11, -9.5) {\footnotesize 68};
\node at (15, -9.5) {miss};
\node[right] at (0,-10.5) {read 100:};
\node at (5, -10.5) {\footnotesize 129};
\node[fill=gray!40] at (7, -10.5) {\footnotesize 100};
\node at (9, -10.5) {\footnotesize 35};
\node at (11, -10.5) {\footnotesize 68};
\node at (15, -10.5) {miss};

\node[right] at (0,-11.5) {read 129:};
\node at (5, -11.5) {\footnotesize 129};
\node at (7, -11.5) {\footnotesize 100};
\node at (9, -11.5) {\footnotesize 35};
\node at (11, -11.5) {\footnotesize 68};
\node at (15, -11.5) {hit};
\node[right] at (0,-12.5) {read 3:};
\node at (5, -12.5) {\footnotesize 129};
\node at (7, -12.5) {\footnotesize 100};
\node[fill=gray!40] at (9, -12.5) {\footnotesize 3};
\node at (11, -12.5) {\footnotesize 68};
\node at (15, -12.5) {miss};
\node[right] at (0,-13.5) {read 35:};
\node at (5, -13.5) {\footnotesize 129};
\node[fill=gray!40] at (7, -13.5) {\footnotesize 35};
\node at (9, -13.5) {\footnotesize 3};
\node at (11, -13.5) {\footnotesize 68};
\node at (15, -13.5) {miss};
\end{tikzpicture}
\vspace{-0.5em}
\caption{Aperiodic memory access of the Fermi L1 data cache. In the figure, the numbers are the data line indices. In the second row, ``read 129" stands for loading the 129th data line, and ``miss" is the memory access status given by the output memory latency array. The highlighted data blocks represent the replaced cache ways according to the output index array when cache misses occur. }
\label{fig:ProcessL1}
\end{figure}

\subsection{L2 Data Cache}


The GTX560Ti, GTX780 and GTX980 report the maximum L2 cache size as 512 KB, 1536 KB and 2048 KB, respectively. Our fine-grained P-chase microbenchmark method is restricted by the shared memory size. At least 64 KB of shared memory is required for a single CTA to store one round of the smallest Fermi L2 cache accesses, much more than our hardware device can offer. However, our fine-grained P-chase can still find the following interesting results.

\begin{enumerate}
  \item The replacement policy of the L2 cache is not LRU, either, because our experimental results show that the memory access processes are aperiodic again.
  \item The L2 cache line size is 32 bytes by observing the memory access pattern of overflowing the cache and visiting array element one by one. The data mapping is sophisticated and not conventional bits-defined, either, since the cache miss pattern is very irregular.
  \item We detect a hardware-level pre-fetching mechanism from the DRAM to the L2 data cache on all three platforms. For example, when we visit an array with uniform stride P-chase, we only observe a long latency for the first data item; the latencies of the following data items all match the L2 cache latency. The pre-fetching size is about 2/3 of the L2 cache size and the pre-fetching is sequential. This is deduced from that if we load an array smaller than 2/3 of the L2 data cache size, there is no cold cache miss patterns.
\end{enumerate}

\vspace{1em}

To summarize, in this section, we study the various GPU caches of three generations of GPUs. We propose a novel fine-grained P-Chase microbenchmark that provides the most detailed measurements. We list the derived parameters of various GPU caches in Table \ref{tab:parameters}. According to our experimental results, the GPU caches are quite different from those of a CPU: they have unequal cache sets and a special replacement policy or data mapping. None of the GPU caches use the traditional bits-defined memory addressing stated in Assumption \ref{def:cond2}. To the best of our knowledge, most of these characteristics have been ignored in previous micro-benchmark GPU studies.


\section{Global Memory}

In CUDA terms, global memory access involves accessing the DRAM, L1 and L2 data caches, TLBs and page tables. It is the most frequently accessed memory space in GPU programming. In this section, we use a series of microbenchmarks to quantitatively study the global memory throughputs and data access latencies on recent GPU platforms.

\subsection{Global Memory Throughput}
\begin{figure*}
  \centering
  \includegraphics[width=0.88\textwidth]{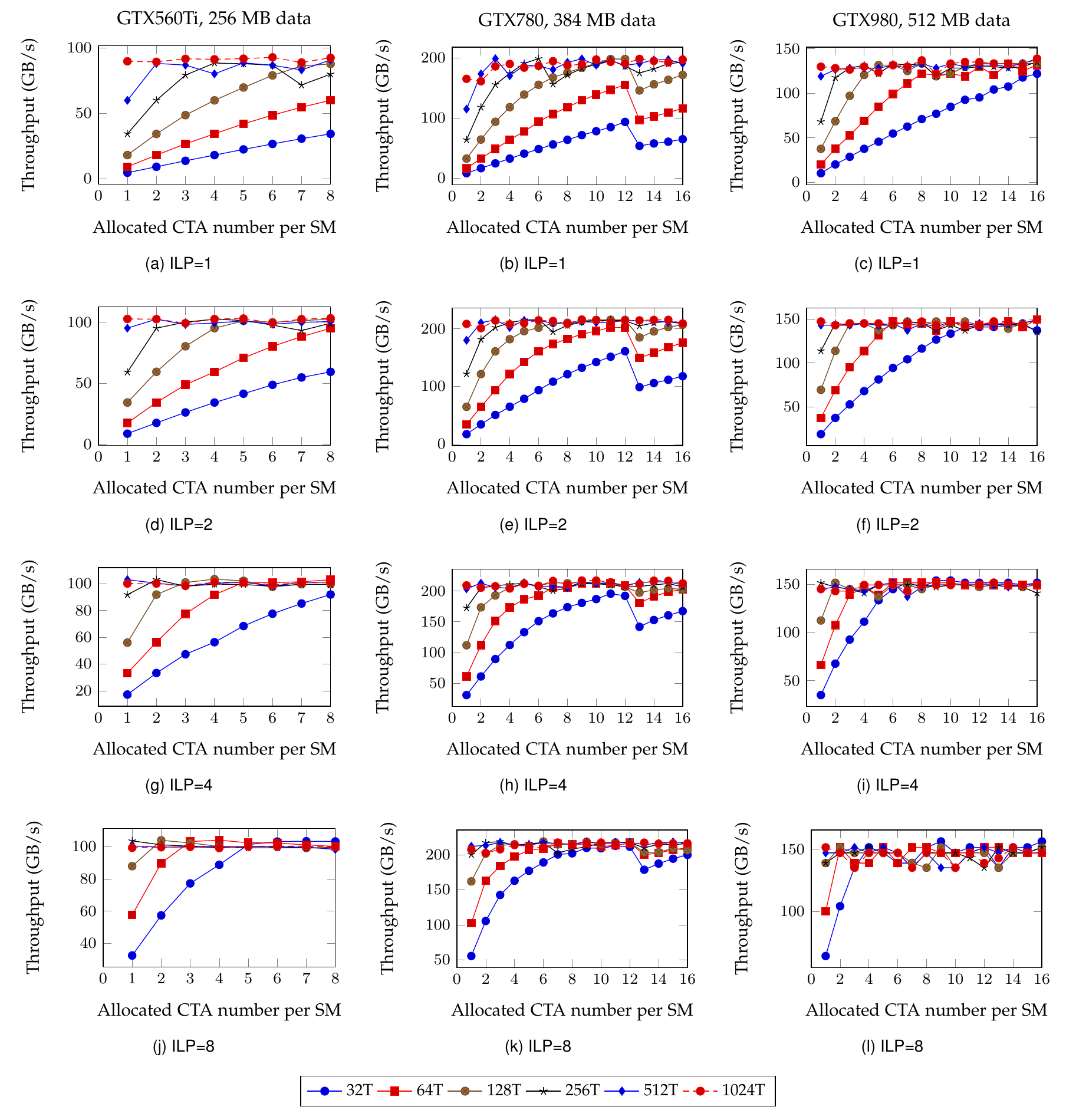}
  \vspace{-1em}
  \caption{Achieved throughput of global memory copy against the number of CTAs, CTA size and ILP. }\label{fig:gbandwidth}
\end{figure*}

Although GPUs are designed with high memory bandwidth, their peak performance can rarely be achieved in reality. The theoretical bandwidth is calculated as $f_{mem}$ * bus width * DDR\_factor, where $f_{mem}$ is the memory frequency, and the DDR\_factor is 4 on all three target platforms. Table \ref{tab:bandwidth} lists the theoretical peak bandwidth and our measured maximum throughput of the three devices.

\begin{table}
\renewcommand{\arraystretch}{1.1}
\centering
\caption{Theoretical and Achieved Bandwidth of Global Memory}
\label{tab:bandwidth}
\begin{tabular}{|c|c|c|c|}
  \hline
  Device & GTX560Ti&GTX780&GTX980 \\ \hline
  $f_{mem}$ (MHz) &1050&1502&1753 \\ \hline
  Bus width (bits) & 256 &384 & 256 \\ \hline
  Theoretical bandwidth (GB/s) & 134.40 & 288.38 & 224.38 \\ \hline
  Maximum throughput (GB/s) & 109.38 & 215.92 & 156.25 \\ \hline
  Efficiency (\%) & 81.38 & 74.87 & 69.64 \\
  \hline
\end{tabular}
\end{table}

The global memory throughput is affected by many factors. According to Little's law, it requires as many memory requests on the fly as possible to fully utilize the bandwidth. We perform a plain memory copy on our three devices with large, fixed amounts of data. We measure the total elapsed time on the CPU. The throughput is calculated as $2 * data size / time$. For each group of experiments, we vary the CTA number, the CTA size (number of threads in each CTA) and the ILP \cite{volkov2010better}. The ILP is defined as the number of 4-byte words that each thread copies at one time. Note that we allocate a number of CTAs to an SM, but these CTAs do not always execute in parallel because the number of activate threads in each SM is limited. Each thread executes the data copying for hundreds of times to ensure there are sufficient memory requests. We plot the achieved throughput in Fig. \ref{fig:gbandwidth}, where $T$ stands for the number of threads per CTA. In general, the throughput converges to its maximum when the ILP/CTA size and the number of CTAs are large. We find that the throughput is limited by the number of active warps: when the size and the number of CTAs are both small, throughput increases almost linearly. The ILP also influences the throughput. Fig. \ref{fig:gbandwidth} shows that for all three devices, the throughput of a larger ILP saturates faster. The GTX560Ti relies on ILP the most, because its SM can launch the fewest warps/CTAs, and a larger ILP helps to handle more memory requests. The GTX780 has the highest throughput as it benefits from the highest bus width, but its convergence speed is the slowest, i.e., it requires the most memory requests to hide the pipeline latency. Considering that such a large amount of parallel memory requests is hardly ever reached in real applications, the higher bus width is somewhat wasteful. This could be part of the reason that NVIDIA reduced the bus width back to 256 bits in Maxwell devices.

\subsection{Global Memory Latency}

In this section, we report the global memory latencies of various data access patterns. The global memory access latency is the whole time accessing a data located in DRAM/L2 or L1 cache, including the latency of page table look-ups. We apply our fine-grained P-chase with a novel self-defined data initialization so that we can collect as many memory latencies as possible in one experiment. We manually set the values of the array elements to create non-uniform stride accesses, rather than executing Listing \ref{list:P-chase_initial}. We are motivated by the convenience of Saavedra1992 method that a single $tvalue$-$s$ graph can show memory latencies of different memory access patterns. Fig. \ref{fig:GIniCompare} illustrates the difference of the data access process between the conventional P-chase and our non-uniform stride fine-grained P-chase.

We measure the global memory latencies with the L1 data cache of the GTX980 and GTX560Ti turned both on and off through the command options. By default, the Maxwell L1 cache is turned off and the Fermi L1 cache is turned on.

Fig. \ref{fig:gAllLatency} shows the global memory latency cycles of six access patterns (noted as P1-P6). In our fine-grained P-chase initialization, we first set a very large $s_1=32$ MB to construct the TLB/page table miss and cache miss (P5\&P6). We then set $s_2=1$ MB to construct the L1 TLB hit but cache miss (P4). After a total of 65 data accesses, 65 data lines are loaded into the cache. We then visit the cached data lines with $s_1$ again for several times, to construct cache hit but TLB miss (P2\&P3). At last, we set $s_3=1$ element and repeatedly load the data in a cache line so that every memory access is a cache hit (P1). The latency values in Fig. \ref{fig:gAllLatency} are based on the average of ten times of experiments. The data cache represents the L1 cache with the GTX980 and GTX560Ti L1 data cache turned on, otherwise it represents the L2 cache. We list some of our findings as follows.

\begin{figure}
\centering
\begin{tikzpicture}[scale=0.35]
    \draw[step=1] (0,0) grid (24, 1);
    \foreach \x in {0,2,4,...,22}
        \draw[fill=red!20] (\x,0) rectangle (\x+1,1);
    \foreach \x in {0,...,23}
        \node(\x) at (\x+0.5,0.5) {\tiny $\x$};
    \foreach \x in {0,2,4,...,20}
        \draw[->] (\x+0.6,1) sin (\x+1.5,1.6) cos (\x+2.4,1);
    \node at (3,2.2) {\scriptsize $s=2$};
\node[below] at (13,0) {\small (a) Uniformed stride};
\end{tikzpicture}
~
\begin{tikzpicture}[scale=0.35]
    \draw[step=1] (0,0) grid (24, 1);

    \draw[fill=red!20] (0,0) rectangle (1,1);
    \draw[fill=red!20] (8,0) rectangle (9,1);
    \draw[->] (0.6,1) sin (4.5,1.6) cos (8.4,1);
    \node at (4.5,2) {\scriptsize $s_1=8$};
    \draw[fill=red!20] (12,0) rectangle (13,1);
    \draw[->] (8.6,0) sin (10.5,-0.6) cos (12.4,0);
    \node at (10.5,-1) {\scriptsize $s_2=4$};

    \draw[fill=red!20] (14,0) rectangle (15,1);
    \draw[->] (12.6,1) sin (13.5,1.6) cos (14.4,1);
    \node at (13.5,2) {\scriptsize $s_3=2$};

    \draw[->] (14.6,0) sin (16.5,-0.6) cos (18.4,0);
    \draw[fill=red!20] (18,0) rectangle (19,1);
    \draw[fill=red!20] (20,0) rectangle (21,1);
    \draw[dashed,->] (18.6,1) sin (19.5,1.6) cos (20.4,1);
    \draw[fill=red!20] (22,0) rectangle (23,1);
    \draw[dashed,->] (20.6,1) sin (21.5,1.6) cos (22.4,1);
    \draw[dashed,->] (22.4,0) sin (20.5,-0.6) cos (18.6,0);
    \foreach \x in {0,...,23}
        \node(\x) at (\x+0.5,0.5) {\tiny $\x$};
    \node[below] at (13,-1.2) {\small (b) Non-uniformed stride};
\end{tikzpicture}
\vspace{-1em}
\caption{Comparison between normal P-chase array access and our non-uniform stride array access. The numbers inside the square blocks are the array indices. The arrows indicate the values of the array elements, for example, the 0th data block pointing to the 2nd block means that we initialize the 0th array element with 2. In Fig. (a), the array is initialized with a single stride $s=2$ that it forms a single memory access pattern: loading every one of two array elements. The measured memory latency is also of this single pattern. In Fig. (b), the array is initialized with various stride, $s_1$, $s_2$ and $s_3$, likewise, we can get the memory latencies of various patterns.}
\label{fig:GIniCompare}
\end{figure}
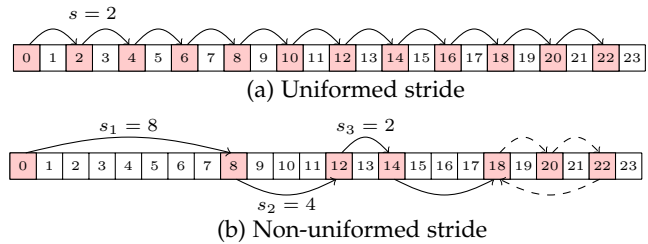

\begin{figure}
\centering{
    \begin{tikzpicture}
    \begin{axis}[small,
        ybar,
        bar width=5pt,
        width=0.49\textwidth,
        height=0.4\textwidth,
        ymin=0,
        ymax=2800,
        ylabel={\footnotesize Memory latency (clock cycles)},
        grid=both,
        symbolic x coords={P1,P2,P3,P4,P5,P6},
	   xtick=data,
        /pgfplots/x grid style={xshift=17pt, draw=brown!80,very thick},
        ybar=0pt,
        legend entries={GTX980-L1on, GTX980-L1off, GTX780, GTX560Ti-L1on, GTX560Ti-L1off},
        legend pos=north west,font=\scriptsize,
    ]
    \addplot[draw=red,fill=red!30,postaction={pattern=north east lines}]
         coordinates{(P1,98) (P2, 0) (P3, 0) (P4,402) (P5,2455) (P6,2756)}[left];
    \addplot[draw=red,fill=red!60]
         coordinates{(P1,230) (P2, 241) (P3, 305) (P4,399) (P5,2477) (P6,2766)};
    \addplot[black,fill=red!30,postaction={pattern=north west lines}]
        coordinates{(P1,230) (P2, 236) (P3, 289) (P4,371) (P5,734) (P6,1000)};
    \addplot[draw=blue,fill=blue!30,postaction={pattern=horizontal lines}]
        coordinates{(P1,116) (P2, 404) (P3, 488) (P4,655) (P5,1259) (P6,0)};
    \addplot[draw=blue,fill=blue!60,postaction={pattern=dots}]
        coordinates{(P1,371) (P2, 398) (P3, 482) (P4,639) (P5,1245) (P6,0)};

    \end{axis}
    \end{tikzpicture}

\vspace{0.5em}
    \renewcommand{\arraystretch}{1.1}
    \footnotesize{
   \centering {
    \begin{tabular}{|c|c|c|c|c|c|c|}
    \hline
    Pattern & P1 & P2 & P3 & P4 & P5 & P6 \\
    \hline
    Data cache & hit & hit & hit & miss & miss & miss \\ \hline
    L1 TLB & hit & miss & miss & hit & miss & miss \\
    L2 TLB & -- & hit & miss & -- & miss & miss \\ \hline
    \hline

    Latency & P1 & P2 & P3 & P4 & P5 & P6 \\
    \hline
    GTX980-L1on & 82 & -- & -- & 385 & 2439 &2740 \\
    GTX980-L1off & 214 & 225 & 289 & 383 & 2461 &2750 \\    \hline
    GTX780 & 198 & 204 & 257 & 339 & 702 & 968 \\ \hline
    GTX560Ti-L1on & 96 & 384 & 468 & 635 & 1239 & -- \\
    GTX560Ti-L1off & 351 & 378 & 462 & 619 & 1225 & -- \\
    \hline
    \end{tabular}}
    }
    }
    \caption{Global memory access latency spectrum.}
    \label{fig:gAllLatency}
\end{figure}

\begin{enumerate}

\item The Maxwell and Kepler devices have a unique memory access pattern (P6) for page table context switching. When a kernel is launched, only memory page entries of 512 MB are activated. If the thread visits an inactivate page entry, the hardware needs a rather long time to switch between page tables. This phenomena is also reported in \cite{meltzer2013micro} as page table ``miss".

\item The Maxwell L1 data cache addressing does not go through the TLBs or page tables. On the GTX980, there is no TLB miss pattern (i.e., P2 and P3) when the L1 data cache is hit. Once the L1 cache is missed, the access latency increases from tens of cycles to hundreds or even thousands of cycles.

\item The TLBs are off-chip. Fig. \ref{fig:gAllLatency} shows that on the GTX560Ti, if the data are cached in L1, the L1 TLB miss penalty is 288 cycles. If data are cached in L2, the L1 TLB miss penalty is 27 cycles. Because the latter penalty is much smaller, we infer that the physical memory locations of the L1 TLB and L2 data cache are close. The physical memory locations of the L1 TLB and L2 TLB are also close, which means that the L1/L2 TLB and L2 data cache are shared off-chip by all SMs.


\item The GTX780 generally has the shortest global memory latencies, almost half that of the Fermi, with an access pattern of P2-P5. By default, the GTX980 has similar latencies to those of the GTX780 for P1-P4. However, for P5 (caused by the cold cache misses), the access latency is about 3.5 times longer than on the Kepler and twice as long as on the Fermi. The page table context switching of the GTX980 is also much more expensive than that of the GTX780.

\end{enumerate}

To summarize, the Maxwell device has long global memory access latencies for cold cache misses and page table context switching. Except for these rare access patterns, its access latency cycles are close to those of the Kepler device. In our experiment, because the GTX980 has higher $f_{mem}$ than the GTX780, it actually offers the shortest global memory access time (P2-P4).

\section{Shared Memory}
The shared memory is designed with high bandwidth and very short memory latency, and each SM has a dedicated shared memory space. In CUDA programming, different CTAs assigned to the same SM have to share the same physical memory space. On the Fermi and Kepler platforms, the shared memory is physically integrated with the L1 cache. On the Maxwell platform, it occupies a separate memory space. Storing data in shared memory is a recognized optimization strategy for GPU-accelerated applications \cite{li2014accelerating,matrixmul_SDK,zhao2014g}. Programmers move the data into and out of shared memory from global memory before and after arithmetic execution, to avoid the frequent occurrence of long global memory access latencies.

In this section, we micro-benchmark the throughput and latency of shared memory. In particular, we discuss the effects of the bank conflict on shared memory access latency. We report a dramatic improvement in performance for the Maxwell device.

\subsection{Shared Memory Throughput}

\begin{table}

\renewcommand{\arraystretch}{1.2}
\centering
\caption{Theoretical and Achieved Throughput of Shared Memory}
    \begin{tabular}{|c|c|c|c|}
            \hline
            Device & GTX560Ti & GTX780 & GTX980 \\ \hline
            $W_{bank}$ (byte/cycle) & 2 & 8 & 4 \\ \hline
            $f_{core}$ (GHz) & 0.950 & 1.006 & 1.279 \\ \hline
            $W_{SM}$ (GB/s) & 60.80 & 257.54 & 163.84 \\ \hline
            $W_{SM}'$ (GB/s) & 34.90 & 83.81 & 137.41 \\ \hline
            Efficiency (\%) & 57.4 & 32.5 & 83.9 \\ \hline
    \end{tabular}
\label{tab:sharedThroughputParameters}
\end{table}

On all three GPU platforms, the shared memory is organized as 32 memory banks \cite{cudacprogrammingguide}. The bank width of the Fermi and Maxwell devices is 4 bytes, while that of the Kepler device is 8 bytes. Each bank has a bandwidth of $W_{bank}$, as shown in Table \ref{tab:sharedThroughputParameters}. The theoretical peak throughput of each SM ($W_{SM}$) is calculated as $f_{core} * W_{bank} * 32$. Our microbenchmark results indicate that although the bandwidth of shared memory is considerable, the real achieved throughput could be much lower. This is most obvious on our Fermi and Kepler devices.

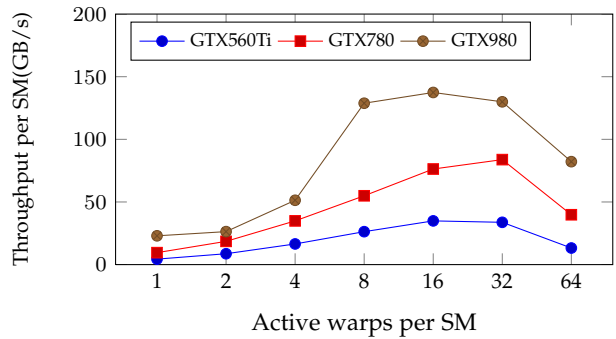
\begin{figure}
\begin{tikzpicture}
    \begin{axis}[small,
        legend columns=-1,
        width=0.45\textwidth,height=0.27\textwidth,
        legend pos=north west, font=\scriptsize,
        symbolic x coords={1,2,4,8,16,32,64},
        ymin=0,ymax=200,
        ylabel={\footnotesize Throughput per SM(GB/s)},
        xlabel={Active warps per SM}
    ]
        \addplot
            table[x=warp_1,y=GTX560Ti] {./SMEM_bandwidth/achievedThroughput.txt};
        \addplot
            table[x=warp_1,y=GTX780] {./SMEM_bandwidth/achievedThroughput.txt};
        \addplot
            table[x=warp_1,y=GTX980] {./SMEM_bandwidth/achievedThroughput.txt};

     \legend{GTX560Ti,GTX780,GTX980}
    \end{axis}
\end{tikzpicture}
\vspace{-1em}
\caption{Achieved shared memory peak throughput per SM.} \label{fig:achievedSMEM}
\end{figure}

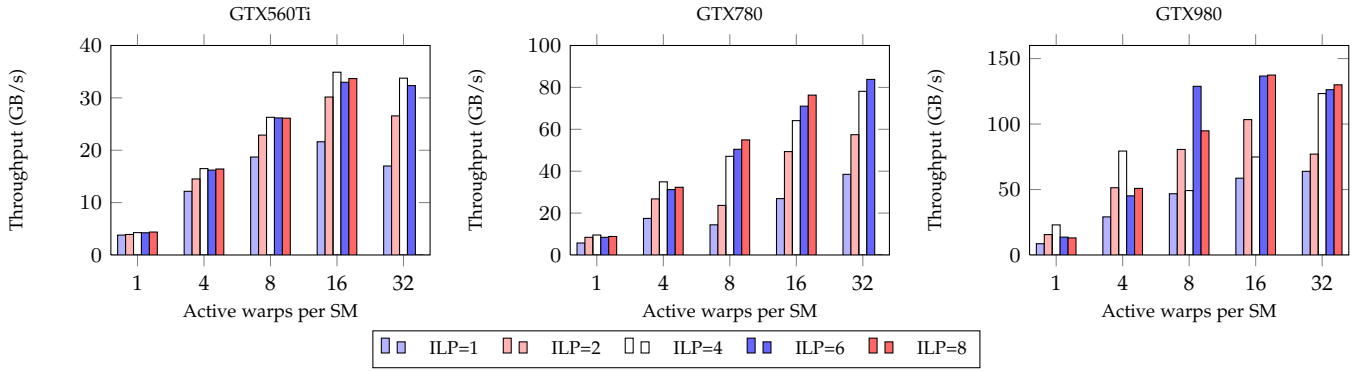
\begin{figure*}
\begin{minipage}[t]{0.32\textwidth}
\begin{tikzpicture}
    \begin{axis}[small,
            legend columns=5,
        legend style={
            column sep=1em,
        },
        legend entries={ILP=1,ILP=2, ILP=4, ILP=6, ILP=8},
        legend to name=named,
        width=1\textwidth,
        height=0.75\textwidth,
        title={\scriptsize GTX560Ti},
        symbolic x coords={1,4,8,16,32},
        ybar=0pt,bar width=3pt,
        ymin=0,ymax=40,
        ylabel={\scriptsize Throughput (GB/s)},
        xlabel={\scriptsize Active warps per SM}
    ]
        \addplot[fill=blue!30,postaction={pattern=north east lines}]
            table[x=WarpsPerSM,y=ILP1] {./SMEM_bandwidth/Fermi.txt};
        \addplot[fill=red!30]
            table[x=WarpsPerSM,y=ILP2] {./SMEM_bandwidth/Fermi.txt};
        \addplot[postaction={pattern=north west lines}]
            table[x=WarpsPerSM,y=ILP4] {./SMEM_bandwidth/Fermi.txt};
        \addplot[fill=blue!60,postaction={pattern=horizontal lines}]
            table[x=WarpsPerSM,y=ILP6] {./SMEM_bandwidth/Fermi.txt};
        \addplot[fill=red!60,postaction={pattern=dots}]
            table[x=WarpsPerSM,y=ILP8] {./SMEM_bandwidth/Fermi.txt};
     \legend{ILP=1,ILP=2, ILP=4, ILP=6, ILP=8}
    \end{axis}
\end{tikzpicture}
\end{minipage}
~
\begin{minipage}[t]{0.32\textwidth}
\begin{tikzpicture}
    \begin{axis}[small,
        width=1\textwidth,
        height=0.75\textwidth,
        title={\scriptsize GTX780},
        symbolic x coords={1,4,8,16,32},
        ybar=0pt,bar width=3pt,
        ymin=0,ymax=100,
        ylabel={\scriptsize Throughput (GB/s)},
        xlabel={\scriptsize Active warps per SM}
    ]
        \addplot[fill=blue!30,postaction={pattern=north east lines}]
            table[x=WarpsPerSM,y=ILP1] {./SMEM_bandwidth/Kepler.txt};
        \addplot[fill=red!30]
            table[x=WarpsPerSM,y=ILP2] {./SMEM_bandwidth/Kepler.txt};
        \addplot[postaction={pattern=north west lines}]
            table[x=WarpsPerSM,y=ILP4] {./SMEM_bandwidth/Kepler.txt};
        \addplot[fill=blue!60,postaction={pattern=horizontal lines}]
            table[x=WarpsPerSM,y=ILP6] {./SMEM_bandwidth/Kepler.txt};
        \addplot[fill=red!60,postaction={pattern=dots}]
            table[x=WarpsPerSM,y=ILP8] {./SMEM_bandwidth/Kepler.txt};

    \end{axis}
\end{tikzpicture}
\end{minipage}
~
\begin{minipage}[t]{0.32\textwidth}
\begin{tikzpicture}
    \begin{axis}[small,
        width=1\textwidth,
        height=0.75\textwidth,
        symbolic x coords={1,4,8,16,32},
        ybar=0pt,bar width=3pt,
        title={\scriptsize GTX980},
        ymin=0,ymax=160,
        ylabel={\scriptsize Throughput (GB/s)},
        xlabel={\scriptsize Active warps per SM}
    ]

        \addplot[fill=blue!30,postaction={pattern=north east lines}]
            table[x=WarpsPerSM,y=ILP1] {./SMEM_bandwidth/Maxwell.txt};
        \addplot[fill=red!30]
            table[x=WarpsPerSM,y=ILP2] {./SMEM_bandwidth/Maxwell.txt};
        \addplot[postaction={pattern=north west lines}]
            table[x=WarpsPerSM,y=ILP4] {./SMEM_bandwidth/Maxwell.txt};
        \addplot[fill=blue!60,postaction={pattern=horizontal lines}]
            table[x=WarpsPerSM,y=ILP6] {./SMEM_bandwidth/Maxwell.txt};
        \addplot[fill=red!60,postaction={pattern=dots}]
            table[x=WarpsPerSM,y=ILP8] {./SMEM_bandwidth/Maxwell.txt};
    \end{axis}
\end{tikzpicture}
\end{minipage}
\centering \scriptsize \ref{named}
\vspace{-1em}
\caption{Shared memory throughput per SM vs. ILP.} \label{fig:SMEMvsILP}
\end{figure*}

The microbenchmark is designed as follows. We copy a number of integers from one shared memory region to another with various grid configurations and ILP levels. Each thread copies ILP of 4-byte data and consumes 8*ILP bytes of shared memory. For each SM, we measure the total elapsed clock cycles with the \emph{\_\_syncthreads}() and \emph{clock}() for all its active warps. The overhead of a pair of \emph{\_\_syncthreads}() and \emph{clock}() is measured as 78, 37, and 36 cycles for Fermi, Kepler, and Maxwell platforms, respectively. The achieved throughput per SM is calculated as 2 * $f_{core}$ * \text{sizeof(int)} * (number of active threads per SM) * ILP / (total latency of each SM). We run the microbenchmark with CTA size = \{32, 64, 128, 256, 512, 1024\}, CTAs per SM =\{1, 2, 3, 4, 5, 6\}, and ILP=\{1, 2, 4, 6, 8\}, subject to the constraint of shared memory size per SM. Usually a large value of ILP results in less active warps per SM. The peak throughput $W_{SM}'$ denotes the respective maximum throughput of the above combinations. Two key factors that affect the throughput are the number of active warps per SM and the ILP level.

We plot the achieved shared memory peak throughput per SM against the number of active warps in Fig. \ref{fig:achievedSMEM}. In general the peak shared memory throughput grows with the increase of active warps, until it reaches some threshold. The peak shared memory throughput of the GTX560Ti occurs when the CTA size = 512, CTAs per SM = 1 and ILP = 4, i.e., 16 active warps per SM. The peak throughput is 34.90 GB/s, which is about 58.7\% of the theoretical bandwidth. The GTX780 reaches its peak throughput when the CTA size = 1024, CTAs per SM = 1 and ILP = 6, i.e., 32 active warps per SM. The peak throughput is 83.81 GB/s, which is only 32.5\% of the theoretical bandwidth. The GTX980 reaches its peak throughput when the CTA size = 256, CTAs per SM = 2 and ILP = 8, i.e., 16 active warps per SM. The peak throughput is 137.41 GB/s, about 83.9\% of the theoretical bandwidth. The Maxwell device shows the best use of its shared memory bandwidth, and the Kepler device shows the worst.

Fig. \ref{fig:SMEMvsILP} shows the achieved shared memory throughputs for different combinations of ILP and number of active warps per SM. Notice that on GTX560Ti and GTX780, when there are 32 active warps, the maximum ILP is 6 due to limited shared memory size. On the GTX560Ti, the achieved throughput grows with the increase of ILP until it reaches 4. On the GTX780, for low SM occupancy (i.e., 1 to 4 active warps), ILP = 4 gives the highest throughput; while for higher SM occupancy (i.e., 8 to 32 active warps), ILP = 6 or 8 give the highest throughput. GTX980 exhibits similar behavior as GTX780: high ILP is required to achieve high throughput for high SM occupancy.

According to Little's Law, we roughly have: number of active warps * ILP = latency cycles * throughput. Applying the latency values in Section 6.2, the GTX780 requires about 94 active warps if ILP = 1, but the Kepler device allows 64 warps at most to be executed concurrently \cite{cudacprogrammingguide}. The gap between the number of required active warps and the number of allowed concurrent warps is particularly obvious on the GTX780. We consider this to be the main reason the achieved throughput of the GTX780 is poor compared with its designed value. For the Maxwell device, due to the significantly reduced access latency, we observe a higher shared memory throughput.

\subsection{Shared Memory Latency}
\renewcommand{\thelstlisting}{\arabic{lstlisting}}
\begin{lstlisting}[caption={Kernel function of shared memory stride access},label={list_shared}]
for ( i=0;i <= iterations; i++ ) {
    data=threadIdx.x*stride;
    if(i==1) sum = 0; //omit cold miss
    start_time = clock();
    repeat64( data=sdata[data];);
        //64 times of stride access
    end_time = clock();
    sum += (end_time - start_time);
}
\end{lstlisting}

We first use the P-chase kernel in Listing \ref{list_shared} with single thread and single CTA to measure the shared memory latencies without bank conflict. The shared memory latencies on Fermi, Kepler and Maxwell devices are 50, 47 and 28 cycles, respectively. However, the shared memory access latency will grow when bank conflicts occur. In this section, we focus on the effect of bank conflicts on shared memory access latency.

The shared memory space is divided into 32 banks. Successive words are allocated to successive banks. If two threads in the same warp access memory spaces in the same bank, a 2-way bank conflict occurs. Listing \ref{list_shared} is also used to measure the shared memory access latency with bank conflicts. Different from the previous case, we launch a warp of threads with a single CTA to access stride memory. We multiply the thread id with an integer, \emph{stride}, to get a shared memory address. We perform the memory access 64 times and record the total time consumption. We then calculate the average memory latency for each memory access.

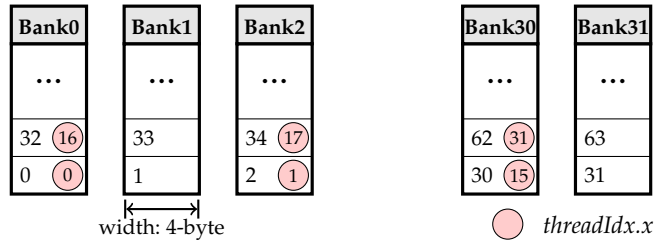
\begin{figure}
\centering
\begin{tikzpicture}[scale=0.5]

    \draw[thick,|<->|] (3,-0.4)--node[below]{\footnotesize width: 4-byte} (5,-0.4);
    \draw[fill=red!20] (13.2, -0.8) circle (0.4);
    \node[right] at (13.9,-0.8){\small \emph{threadIdx.x}};
    \foreach \a in {0,1,2,4,5}{
        \draw[very thick] (\a*3,0) rectangle (\a*3+2,4);
        \draw (\a*3,1) -- (\a*3+2,1) ; \draw (\a*3,2) -- (\a*3+2,2) ;
        \node at (\a*3+1,3) {\Large ...};
    }
    \foreach \x in {0,1,2}{
        \draw[very thick, fill=gray!20] (\x*3,4) rectangle (\x*3+2,5);
        \node[above] at (\x*3+1,4) {\footnotesize \textbf {Bank\x}};
        \foreach \y[evaluate={\i=int(32*\y+\x)}] in {0,1}{
            \node[right] at (\x*3,\y+0.5) {\footnotesize \i};
        }
    }
    \foreach \x[evaluate={\j=int(\x+26)}] in {4,5}{
        \draw[very thick, fill=gray!20] (\x*3,4) rectangle (\x*3+2,5);
        \node[above] at (\x*3+1,4) {\footnotesize \textbf {Bank\j}};
        \foreach \y[evaluate={\i=int(32*\y+\x+26)}] in {0,1}{
            \node[right] at (\x*3,\y+0.5) {\footnotesize \i};
        }
    }

     \foreach \x[evaluate={\i=int(\x+16)}] in {0,1}{
        \draw[fill=red!20] (6*\x+1.5, 0.5) circle (0.4);
        \draw[fill=red!20] (6*\x+1.5, 1.5) circle (0.4);
        \node at (6*\x+1.5, 0.5) {\scriptsize \x};
        \node at (6*\x+1.5, 1.5) {\scriptsize \i};
    }
    \draw[fill=red!20] (13.5, 0.5) circle (0.4);
    \draw[fill=red!20] (13.5, 1.5) circle (0.4);
        \node at (13.5, 0.5) {\scriptsize 15};
        \node at (13.5, 1.5) {\scriptsize 31};
\end{tikzpicture}
\vspace{-1em}
\caption{2-way shared memory bank conflict (stride=2).}
\label{fig:fermi_shared_demo}
\end{figure}

Fig. \ref{fig:fermi_shared_demo} illustrates a 2-way bank conflict caused by stride memory access on the Fermi architecture. For example, word 0 and word 32 are mapped onto the same bank. If the stride is 2, threads 0 and 16 will visit words 0 and 32, respectively, which causes a 2-way bank conflict. The number of potential bank conflicts equals the greatest common divisor of the stride number and 32. There is no bank conflict for odd strides. Fermi and Maxwell devices have the same number of potential bank conflicts because they have the same architecture.

\begin{figure*}
\centering
    \begin{tikzpicture}[scale=0.49]
    \draw[fill=red!20] (30, -0.7) circle (0.4);
    \node[right] at (30.5,-0.7){\small \emph{threadIdx.x}};
    \draw[thick,|<->|] (2,-0.4)--node[below]{\footnotesize width: 4-byte} (4,-0.4);
    \foreach \a in{0,1,2,3,4,6}{
        \draw[very thick] (5*\a,0) rectangle (5*\a+4,5);
        \draw[very thick, fill=gray!20] (\a*5,5) rectangle (5*\a+4,6);
        \draw (5*\a+2,0)--(5*\a+2,5);
        \draw (5*\a,1)--(5*\a+4,1);\draw (5*\a,2)--(5*\a+4,2);\draw (5*\a,3)--(5*\a+4,3);
        \node at (\a*5+1,4) {\Large ...}; \node at (\a*5+3,4) {\Large ...};
    }
    \foreach \x in {0,1,2,3,4}{
    \node[above] at (\x*5+2,5) {\footnotesize \textbf {Bank\x}};
        \foreach \y[evaluate={\i=int(64*\y+\x)},evaluate={\j=int(64*\y+\x+32)}] in {0,1,2}{
            \node[right] at (\x*5,\y+0.5) {\footnotesize \i};
            \node[right] at (\x*5+2,\y+0.5) {\footnotesize \j};
        }
    }
    \node[above] at (32,5) {\footnotesize \textbf {Bank31}};
    \foreach \y[evaluate={\i=int(64*\y+31)},evaluate={\j=int(64*\y+63)}] in {0,1,2}{
            \node[right] at (30,\y+0.5) {\footnotesize \i};
            \node[right] at (32,\y+0.5) {\footnotesize \j};
        }
    \draw[fill=red!20] (1.5, 0.5) circle (0.4); \node at (1.5, 0.5) {\scriptsize 0};
    \draw[fill=red!20] (3.5, 1.5) circle (0.4); \node at (3.5, 1.5) {\scriptsize 16};
    \draw[fill=red!20] (11.5, 1.5) circle (0.4); \node at (11.5, 1.5) {\scriptsize 11};
    \draw[fill=red!20] (13.5, 2.5) circle (0.4); \node at (13.5, 2.5) {\scriptsize 27};
    \draw[fill=red!20] (21.5, 2.5) circle (0.4); \node at (21.5, 2.5) {\scriptsize 22};
    \draw[fill=red!20] (23.5, 0.5) circle (0.4); \node at (23.5, 0.5) {\scriptsize 6};

    \node at (17,-1.2) {(a) 4-byte mode, 2-way bank conflict. };
    \end{tikzpicture}

\hspace{0.1em}

\centering
    \begin{tikzpicture}[scale=0.49]
    \draw[fill=red!20] (30, -0.7) circle (0.4);
    \node[right] at (30.5,-0.7){\small \emph{threadIdx.x}};
        \draw[thick,|<->|] (0,-0.4)--node[below]{\footnotesize width: 8-byte} (4,-0.4);
    \foreach \a in{0,1,2,3,4,6}{
        \draw[very thick] (5*\a,0) rectangle (5*\a+4,5);
        \draw[very thick, fill=gray!20] (\a*5,5) rectangle (5*\a+4,6);
        \draw (5*\a+2,0)--(5*\a+2,5);
        \draw (5*\a,1)--(5*\a+4,1);\draw (5*\a,2)--(5*\a+4,2);\draw (5*\a,3)--(5*\a+4,3);
        \node at (\a*5+1,4) {\Large ...}; \node at (\a*5+3,4) {\Large ...};
    }
    \foreach \x in {0,1,2,3,4}{
    \node[above] at (\x*5+2,5) {\footnotesize \textbf {Bank\x}};
        \foreach \y[evaluate={\i=int(64*\y+\x*2)},evaluate={\j=int(\i+1)}] in {0,1,2}{
            \node[right] at (\x*5,\y+0.5) {\footnotesize \i};
            \node[right] at (\x*5+2,\y+0.5) {\footnotesize \j};
        }
    }
    \node[above] at (32,5) {\footnotesize \textbf {Bank31}};
    \foreach \y[evaluate={\i=int(64*\y+62)},evaluate={\j=int(\i+1)}] in {0,1,2}{
            \node[right] at (30,\y+0.5) {\footnotesize \i};
            \node[right] at (32,\y+0.5) {\footnotesize \j};
        }

    \draw[fill=red!20] (1.5, 0.5) circle (0.4); \node at (1.5, 0.5) {\scriptsize 0};
    \draw[fill=red!20] (16.5, 0.5) circle (0.4); \node at (16.5, 0.5) {\scriptsize 1};
    \draw[fill=red!20] (6.5, 1.5) circle (0.4); \node at (6.5, 1.5) {\scriptsize 11};
    \draw[fill=red!20] (21.5, 1.5) circle (0.4); \node at (21.5, 1.5) {\scriptsize 12};
    \draw[fill=red!20] (31.5, 1.5) circle (0.4); \node at (31.5, 1.5) {\scriptsize 21};
    \draw[fill=red!20] (11.5, 2.5) circle (0.4); \node at (11.5, 2.5) {\scriptsize 22};

    \node at (17,-1.2) {(b) 8-byte mode, no bank conflict.};
    \end{tikzpicture}
\vspace{-1em}
\caption{Kepler shared memory bank conflict (stride = 6).}
\label{fig:kepler_stride6_sm}

\vspace{1em}

\begin{tikzpicture}[scale=0.95]
        \begin{axis}[
            ybar=0pt,
            bar width=5pt,
            legend pos=north west,font=\scriptsize,
            width=1\textwidth, height=0.28\textwidth,
            xlabel={Stride},
            ylabel={Memory latency (clock cycles)},
            xmin=-1, xmax=65,
            ymin=0, ymax=500,
            xtick={0,2,...,64},
            ytick={0,100,...,500},
            xmajorgrids=true,
            ymajorgrids=true
        ]
        \addplot[fill=red!20]
            table[x=stride,y=Kepler_4byte] {./data/shared.txt};
        \addplot[fill=red!60,postaction={pattern=north east lines}]
            table[x=stride,y=Kepler_8byte] {./data/shared.txt};
        \legend{4-byte mode, 8-byte mode};
        \end{axis}
        \end{tikzpicture}
\vspace{-1em}
  \caption{Latency of Kepler Shared Memory with bank conflict: 4-byte mode v.s. 8-byte mode.}
  \label{fig:kepler_shared_latency}
\end{figure*}

Kepler outperforms Fermi in terms of avoiding shared memory bank conflicts by doubling the bank width \cite{micikevicius2012gpu}. The bank width of Kepler device is 8 bytes, yet it offers two configurable modes to programmers: 4-byte mode and 8-byte mode. In the 8-byte mode, 64 successive integers are mapped onto 32 successive banks, whereas in the 4-byte mode, 32 successive integers are mapped onto 32 successive banks. Fig. \ref{fig:kepler_stride6_sm} illustrates the data mapping of the two modes. A bank conflict only occurs when two or more threads access different bank rows.
Fig. \ref{fig:kepler_shared_latency} shows the Kepler shared memory latencies with even strides for the 4-byte and 8-byte modes. When the stride is 2, there is no bank conflict in either mode, whereas there is a 2-way bank conflict on Fermi. When the stride is 4, both modes show a 2-way bank conflict. When the stride is 6 (Fig. \ref{fig:kepler_stride6_sm}), there is a 2-way bank conflict for the 4-byte mode but no bank conflict for the 8-byte mode. For the 4-byte mode, half of the shared memory banks are visited. Thread $i$ and thread $i+16$ access separate rows in the same bank ($i=0,...,15$). For the 8-byte mode, 32 threads visit 32 different banks with no conflict. Similarly, the 8-byte mode is superior to the 4-byte mode for other even strides if their number is not to the power of two.

\begin{table}

\renewcommand{\arraystretch}{1.1}
\centering
\caption{Shared Memory Access Latency with Bank Conflicts}
    \begin{tabular}{|c|c|c|c|c|c|}
            \hline
            Bank conflict& 2-way & 4-way &8-way&16-way&32-way \\ \hline
            GTX980  & 30& 34&42&58&90 \\ \hline
            GTX780  & 82 & 96 & 158 &257 &484\\ \hline
            GTX560Ti &87 &162 &311&611&1209 \\ \hline
    \end{tabular}
\label{tab:smemLatencyBankConflict}
\end{table}

We list our measured shared memory access latencies according to the number of potential bank conflicts in Table \ref{tab:smemLatencyBankConflict}. The memory access latency increases almost linearly with the number of potential bank conflicts. This confirms that the data access instructions are sequentially executed in case of a bank conflict. For the Fermi and Kepler devices, where there is a 32-way bank conflict, it takes much longer to access shared memory than regular global memory (TLB hit, cache miss). Surprisingly, the effect of a bank conflict on shared memory access latency on the Maxwell device is mild. Even the longest shared memory access latency is still at the same level as L1 data cache latency.

In summary, although the shared memory has very short access latency, it can be rather long if there are many ways of bank conflicts. This is most obvious on the Fermi hardware. The Kepler device tries to solve it by doubling the bank width of shared memory. Compared with the Fermi, the Kepler's 4-byte mode shared memory halves the chance of bank conflict, and the 8-byte mode reduces it further. However, we also find that the Kepler's shared memory is inefficient in terms of throughput. The Maxwell device has the best shared memory performance. With the same architecture as the Fermi device, the Maxwell hardware shows a 2x size, 2x memory access speedup and achieves the highest throughput. Most importantly, the Maxwell device's shared memory has been optimized to avoid the long latency caused by bank conflicts. As many GPU-accelerated applications rely on shared memory performance, this improvement certainly leads to faster and more efficient GPU computations.

\section{Conclusions}

In this study, we microbenchmarked the cache characteristics, memory throughput, and memory latencies of three recent generations of NVIDIA GPUs: Fermi, Kepler and Maxwell. We perceive an evolution of the NVIDIA GPU memory hierarchy. The memory capacity is significantly enhanced in both Kepler and Maxwell as compared with Fermi. The Kepler device is performance-oriented and incorporates several aggressive elements in its design, such as increasing the bus width of DRAM and doubling the bank width of shared memory. These designs have some side-effects. The theoretical bandwidths of both global memory and shared memory are difficult to saturate, and hardware resources are imbalanced with a low utilization rate. The Maxwell device has a more efficient and conservative design. It has a reduced bus width and bank width, and the on-chip cache architectures are adjusted, including doubling the shared memory size and the read-only data cache size. Furthermore, it sharply decreases the shared memory latency caused under bank conflicts. With its optimized memory hierarchy, the Maxwell device not only retains good performance but is also more economical.

\section*{Acknowledgement}
We thank the anonymous reviewers for their valuable comments. This work is partially supported by Hong Kong GRF grant HKBU 210412, HKBU FRG2/14-15/059, Shenzhen Basic Research Grant SCI-2015-SZTIC-002.


\bibliographystyle{IEEEtran}
\bibliography{citations}

\begin{thebibliography}{10}
\providecommand{\url}[1]{#1}
\csname url@samestyle\endcsname
\providecommand{\newblock}{\relax}
\providecommand{\bibinfo}[2]{#2}
\providecommand{\BIBentrySTDinterwordspacing}{\spaceskip=0pt\relax}
\providecommand{\BIBentryALTinterwordstretchfactor}{4}
\providecommand{\BIBentryALTinterwordspacing}{\spaceskip=\fontdimen2\font plus
\BIBentryALTinterwordstretchfactor\fontdimen3\font minus
  \fontdimen4\font\relax}
\providecommand{\BIBforeignlanguage}[2]{{%
\expandafter\ifx\csname l@#1\endcsname\relax
\typeout{** WARNING: IEEEtran.bst: No hyphenation pattern has been}%
\typeout{** loaded for the language `#1'. Using the pattern for}%
\typeout{** the default language instead.}%
\else
\language=\csname l@#1\endcsname
\fi
#2}}
\providecommand{\BIBdecl}{\relax}
\BIBdecl

\bibitem{nickolls2010gpu}
J.~Nickolls and W.~J. Dally, ``The {GPU} computing era,'' \emph{IEEE Micro},
  vol.~30, no.~2, pp. 56--69, 2010.

\bibitem{Hwu20142574}
W.~mei Hwu, ``What is ahead for parallel computing,'' \emph{Journal of Parallel
  and Distributed Computing}, vol.~74, no.~7, pp. 2574--2581, 2014.

\bibitem{keckler2011gpus}
S.~W. Keckler, W.~J. Dally, B.~Khailany, M.~Garland, and D.~Glasco, ``{GPUs}
  and the future of parallel computing,'' \emph{IEEE Micro}, vol.~31, no.~5,
  pp. 0007--17, 2011.

\bibitem{zhao2014g}
K.~Zhao and X.~Chu, ``{G-BLASTN}: accelerating nucleotide alignment by graphics
  processors,'' \emph{Bioinformatics}, vol.~30, no.~10, pp. 1384--91, 2014.

\bibitem{li2014accelerating}
Y.~Li, H.~Chi, L.~Xia, and X.~Chu, ``Accelerating the scoring module of mass
  spectrometry-based peptide identification using {GPUs},'' \emph{BMC
  bioinformatics}, vol.~15, no.~1, pp. 1--11, 2014.

\bibitem{ryoo2008optimization}
S.~Ryoo, C.~I. Rodrigues, S.~S. Baghsorkhi, S.~S. Stone, D.~B. Kirk, and
  W.-m.~W. Hwu, ``Optimization principles and application performance
  evaluation of a multithreaded {GPU} using {CUDA},'' in \emph{Proc. of the
  13th ACM SIGPLAN Symposium on Principles and Practice of Parallel
  Programming}.\hskip 1em plus 0.5em minus 0.4em\relax ACM, 2008, pp. 73--82.

\bibitem{MaxwellWhitepaper}
\emph{{{NVIDIA GeForce GTX} 980 Whitepaper}}, NVIDIA Corporation, 2014.

\bibitem{micikevicius20093d}
P.~Micikevicius, ``{3D} finite difference computation on {GPUs} using {CUDA},''
  in \emph{Proc. of 2nd Workshop on General Purpose Processing on Graphics
  Processing Units}.\hskip 1em plus 0.5em minus 0.4em\relax ACM, 2009, pp.
  79--84.

\bibitem{matrixmul_SDK}
NVIDIA, ``{matrixMul},'' \emph{CUDA SDK 6.5}, 2014.

\bibitem{matrixmulCUBLAS_SDK}
------, ``{matrixMulCUBLAS},'' \emph{CUDA SDK 6.5}, 2014.

\bibitem{fermiwhitepaper}
\emph{Fermi Whitepaper}, NVIDIA Corporation, 2009.

\bibitem{keplerwhitepaper}
\emph{Kepler {GK110} Whitepaper}, NVIDIA Corporation, 2012.

\bibitem{keplertuningguide}
\emph{Tuning {CUDA} Applications for Kepler}, NVIDIA Corporation, 2013.

\bibitem{MaxwellTuning}
\emph{Tuning {CUDA} Applications for Maxwell}, NVIDIA Corporation, 2014.

\bibitem{cudacprogrammingguide}
\emph{{CUDA C} Programming Guide - v7.5}, NVIDIA Corporation, 2015.

\bibitem{cudabestguide}
\emph{{CUDA C} Best Practices Guide - v7.5}, NVIDIA Corporation, 2015.

\bibitem{volkov2008benchmarking}
V.~Volkov and J.~W. Demmel, ``Benchmarking {GPUs} to tune dense linear
  algebra,'' in \emph{Proc. of the 2008 ACM/IEEE Conference on Supercomputing},
  no.~31.\hskip 1em plus 0.5em minus 0.4em\relax IEEE Press, 2008.

\bibitem{papadopoulou2009micro}
M.~Papadopoulou, M.~Sadooghi-Alvandi, and H.~Wong, ``Micro-benchmarking the
  {GT200 GPU},'' \emph{Computer Group, ECE, University of Toronto, Tech. Rep},
  2009.

\bibitem{wong2010demystifying}
H.~Wong, M.-M. Papadopoulou, M.~Sadooghi-Alvandi, and A.~Moshovos,
  ``Demystifying {GPU} microarchitecture through microbenchmarking,'' in
  \emph{Proc. of Performance Analysis of Systems and Software (ISPASS), 2010
  IEEE International Symposium on}.\hskip 1em plus 0.5em minus 0.4em\relax
  IEEE, 2010, pp. 235--246.

\bibitem{zhang2011quantitative}
Y.~Zhang and J.~D. Owens, ``A quantitative performance analysis model for {GPU}
  architectures,'' in \emph{Proc. of High Performance Computer Architecture
  (HPCA), 2011 IEEE 17th International Symposium on}.\hskip 1em plus 0.5em
  minus 0.4em\relax IEEE, 2011, pp. 382--393.

\bibitem{baghsorkhi2012efficient}
S.~S. Baghsorkhi, I.~Gelado, M.~Delahaye, and W.-m.~W. Hwu, ``Efficient
  performance evaluation of memory hierarchy for highly multithreaded graphics
  processors,'' in \emph{ACM SIGPLAN Notices}, vol.~47, no.~8.\hskip 1em plus
  0.5em minus 0.4em\relax ACM, 2012, pp. 23--34.

\bibitem{meltzer2013micro}
R.~Meltzer, C.~Zeng, and C.~Cecka, ``Micro-benchmarking the {C2070},'' 2013,
  poster presented at GPU Technology Conference, March 18-21, San Jose,
  California.

\bibitem{saavedra1992cpu}
R.~H. Saavedra, ``{CPU} performance evaluation and execution time prediction
  using {Narrow} spectrum benchmarking,'' Ph.D. dissertation, EECS Department,
  University of California, Berkeley, Feb 1992.

\bibitem{saavedra1995measuring}
R.~H. Saavedra and A.~J. Smith, ``Measuring cache and {TLB} performance and
  their effect on benchmark runtimes,'' \emph{Computers, IEEE Transactions on},
  vol.~44, no.~10, pp. 1223--1235, 1995.

\bibitem{mei2014benchmarking}
X.~Mei, K.~Zhao, C.~Liu, and X.~Chu, ``Benchmarking the memory hierarchy of
  modern {GPUs},'' in \emph{Proc. of Network and Parallel Computing, 2014 IFIP
  11th International Conference on}, 2014, pp. 144--156.

\bibitem{lal2014gpgpu}
S.~Lal, J.~Lucas, M.~Andersch, M.~Alvarez-Mesa, A.~Elhossini, and B.~Juurlink,
  ``{GPGPU} workload characteristics and performance analysis,'' in \emph{Proc.
  of Embedded Computer Systems: Architectures, Modeling, and Simulation (SAMOS
  XIV), 2014 International Conference on}.\hskip 1em plus 0.5em minus
  0.4em\relax IEEE, 2014, pp. 115--124.

\bibitem{jia2012characterizing}
W.~Jia, K.~A. Shaw, and M.~Martonosi, ``Characterizing and improving the use of
  demand-fetched caches in {GPUs},'' in \emph{Proc. of the 26th ACM
  International Conference on Supercomputing}.\hskip 1em plus 0.5em minus
  0.4em\relax ACM, 2012, pp. 15--24.

\bibitem{xie2013efficient}
X.~Xie, Y.~Liang, G.~Sun, and D.~Chen, ``An efficient compiler framework for
  cache bypassing on {GPUs},'' in \emph{Proc. of Computer-Aided Design (ICCAD),
  2013 IEEE/ACM International Conference on}.\hskip 1em plus 0.5em minus
  0.4em\relax IEEE, 2013, pp. 516--523.

\bibitem{jang2011exploiting}
B.~Jang, D.~Schaa, P.~Mistry, and D.~Kaeli, ``Exploiting memory access patterns
  to improve memory performance in data-parallel architectures,''
  \emph{Parallel and Distributed Systems, IEEE Transactions on}, vol.~22,
  no.~1, pp. 105--118, 2011.

\bibitem{Che2011Dymaxion}
S.~Che, J.~W. Sheaffer, and K.~Skadron, ``Dymaxion: Optimizing memory access
  patterns for heterogeneous systems,'' in \emph{Proc. of 2011 International
  Conference for High Performance Computing, Networking, Storage and Analysis},
  2011, pp. 13:1--13:11.

\bibitem{sung2010data}
I.-J. Sung, J.~A. Stratton, and W.-M.~W. Hwu, ``Data layout transformation
  exploiting memory-level parallelism in structured grid many-core
  applications,'' in \emph{Proc. of the 19th International Conference on
  Parallel Architectures and Compilation Techniques}.\hskip 1em plus 0.5em
  minus 0.4em\relax ACM, 2010, pp. 513--522.

\bibitem{li2015}
C.~Li, S.~L. Song, H.~Dai, A.~Sidelnik, S.~K.~S. Hari, and H.~Zhou,
  ``Locality-driven dynamic {GPU} cache bypassing,'' in \emph{Proceedings of
  the 29th ACM on International Conference on Supercomputing}.\hskip 1em plus
  0.5em minus 0.4em\relax ACM, 2015, pp. 67--77.

\bibitem{mcvoy1996lmbench}
L.~W. McVoy, C.~Staelin \emph{et~al.}, ``lmbench: Portable tools for
  performance analysis.'' in \emph{Proc. of USENIX Annual Technical
  Conference}.\hskip 1em plus 0.5em minus 0.4em\relax San Diego, CA, USA, 1996,
  pp. 279--294.

\bibitem{duchateau2008p-ray}
A.~X. Duchateau, A.~Sidelnik, M.~J. Garzar{\'a}n, and D.~Padua, ``{P-ray: A
  software suite for multi-core architecture characterization},'' in
  \emph{Proc. of Languages and Compilers for Parallel Computing, 21th
  International Workshop on}.\hskip 1em plus 0.5em minus 0.4em\relax Springer,
  2008, pp. 187--201.

\bibitem{hakura1997design}
Z.~S. Hakura and A.~Gupta, ``The design and analysis of a cache architecture
  for texture mapping,'' \emph{ACM SIGARCH Computer Architecture News},
  vol.~25, no.~2, pp. 108--120, 1997.

\bibitem{volkov2010better}
V.~Volkov, ``Better performance at lower occupancy,'' in \emph{the 1st GPU
  Technology Conference}.\hskip 1em plus 0.5em minus 0.4em\relax San Jose, CA,
  USA, 2010.

\bibitem{micikevicius2012gpu}
P.~Micikevicius, ``{GPU} performance analysis and optimization,'' in \emph{the
  3rd GPU Technology Conference}.\hskip 1em plus 0.5em minus 0.4em\relax San
  Jose, CA, USA, 2012.

\end{thebibliography}
\begin{IEEEbiography}
[{\includegraphics[width=1in,height=1.25in,clip,keepaspectratio]{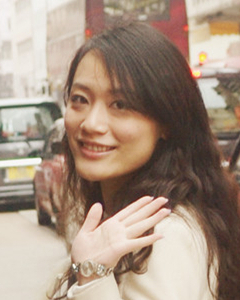}}]
{Xinxin Mei}
received the B.E. degree in electronic information engineering from the University of Science and Technology of China, P.R.C., in 2010. She is currently a Ph.D. student in the Department of Computer Science, Hong Kong Baptist University. Her research interests include distributed and parallel computing and GPU-accelerated parallel partial differential equation solvers.
\end{IEEEbiography}

\begin{IEEEbiography}
[{\includegraphics[width=1in,height=1.25in,clip,keepaspectratio]
{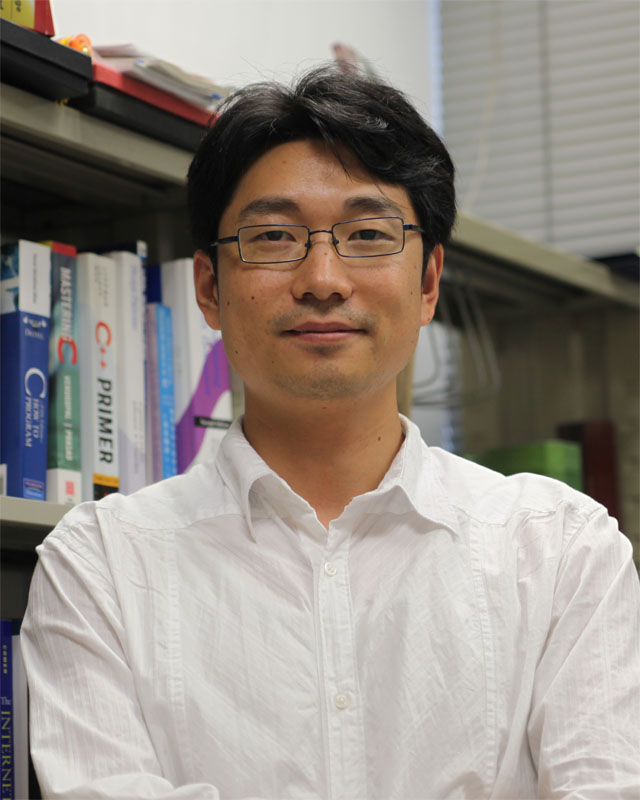}}]{Xiaowen Chu}
received the B.E. degree in computer science from Tsinghua University, P.R. China, in 1999, and the Ph.D. degree in computer science from the Hong Kong University of Science and Technology in 2003. Currently, he is an associate professor in the Department of Computer Science, Hong Kong Baptist University. His research interests include distributed and parallel computing and wireless networks. He is serving as an Associate Editor of IEEE Access and IEEE Internet of Things Journal. He is a senior member of the IEEE.
\end{IEEEbiography}

\end{document}